\documentclass[aps,prb,twocolumn,superscriptaddress,showpacs%
]{revtex4-1}	%
\usepackage[ngerman,english]{babel} %
\usepackage[utf8]{inputenc}

\usepackage{xcolor,soul}

\usepackage{amsmath,amssymb} %

\usepackage[pdfstartview=FitH,      %
            breaklinks=true,        %
            bookmarksopen=true,     %
            bookmarksnumbered=true  %
            ]{hyperref}				%
            
\usepackage{wrapfig}
\usepackage{subfig}   

\usepackage{adjustbox} 		%
\usepackage{upgreek}		%
\usepackage{xargs}                      %

\usepackage[colorinlistoftodos,prependcaption,textsize=tiny]{todonotes}
\newcommandx{\unsure}[2][1=]{\todo[linecolor=red,backgroundcolor=red!25,bordercolor=red,#1]{#2}}
\newcommandx{\change}[2][1=]{\todo[linecolor=blue,backgroundcolor=blue!25,bordercolor=blue,#1]{#2}}
\newcommandx{\info}[2][1=]{\todo[linecolor=OliveGreen,backgroundcolor=OliveGreen!25,bordercolor=OliveGreen,#1]{#2}}
\newcommandx{\improvement}[2][1=]{\todo[linecolor=Plum,backgroundcolor=Plum!25,bordercolor=Plum,#1]{#2}}
\newcommandx{\thiswillnotshow}[2][1=]{\todo[disable,#1]{#2}}

\usepackage{nicefrac}
\usepackage{braket}
\usepackage[mathscr]{eucal}	
\usepackage{bbm}
\usepackage{floatrow}    %

\begin{document}

\title{Stabilization of phase noise in spin torque nano oscillators by a phase locked loop }

\author{Steffen Wittrock}\altaffiliation[Present address: ]{Max Born Institute For Nonlinear Optics \& Short Pulse Spectroscopy, Max-Born-Str. 2A, 12489 Berlin, Germany}\email[Email: ]{steffen.wittrock@mbi-berlin.de}\affiliation{Unit\'{e} Mixte de Physique CNRS, Thales, Univ. Paris-Saclay, 1 Avenue Augustin Fresnel, 91767, Palaiseau, France }
\author{Martin Kreißig}\affiliation{Chair for Circuit Design and Network Theory, Technische Universität Dresden, 01062 Dresden, Germany}
\author{Bertrand Lacoste}\affiliation{International Iberian Nanotechnology Laboratory (INL), 471531 Braga, Portugal}
\author{Artem Litvinenko}\affiliation{Univ. Grenoble Alpes, CEA, INAC-SPINTEC, CNRS, SPINTEC, 38000 Grenoble, France}
\author{Philippe Talatchian}\affiliation{Univ. Grenoble Alpes, CEA, INAC-SPINTEC, CNRS, SPINTEC, 38000 Grenoble, France}
\author{Florian Protze}\affiliation{Chair for Circuit Design and Network Theory, Technische Universität Dresden, 01062 Dresden, Germany}
\author{Frank Ellinger}\affiliation{Chair for Circuit Design and Network Theory, Technische Universität Dresden, 01062 Dresden, Germany}

\author{Ricardo Ferreira}\affiliation{International Iberian Nanotechnology Laboratory (INL), 471531 Braga, Portugal}
\author{Romain Lebrun}\affiliation{Unit\'{e} Mixte de Physique CNRS, Thales, Univ. Paris-Saclay, 1 Avenue Augustin Fresnel, 91767 Palaiseau, France}
 \author{Paolo Bortolotti}\affiliation{Unit\'{e} Mixte de Physique CNRS, Thales, Univ. Paris-Saclay, 1 Avenue Augustin Fresnel, 91767 Palaiseau, France}
  \author{Liliana Buda-Prejbeanu}\affiliation{Univ. Grenoble Alpes, CEA, INAC-SPINTEC, CNRS, SPINTEC, 38000 Grenoble, France}
 \author{Ursula Ebels}\email[Email: ]{ursula.ebels@cea.fr}\affiliation{Univ. Grenoble Alpes, CEA, INAC-SPINTEC, CNRS, SPINTEC, 38000 Grenoble, France}

 \author{Vincent Cros}\email[Email: ]{vincent.cros@cnrs-thales.fr}\affiliation{Unit\'{e} Mixte de Physique CNRS, Thales, Univ. Paris-Saclay, 1 Avenue Augustin Fresnel, 91767 Palaiseau, France}

\date{\today}

\begin{abstract}

The main limitation in order to exploit spin torque nano-oscillators (STNOs) in various potential applications is their large phase noise. 
In this work, we demonstrate its efficient reduction  by a highly reconfigurable, compact, specifically on-chip designed PLL based on custom integrated circuits. 
First, we thoroughly study the parameter space of the PLL+STNO system experimentally. 
Second, we present a theory which describes the locking of a STNO to an external signal in a general sense. 
In our developed theory, we do not restrict ourselves to the case of a perfect phase locking but also consider phase slips and the corresponding low offset frequency $1/f^2$ noise, so far the main drawback in such systems. 
Combining experiment and theory allows us to reveal complex parameter dependences of the system's phase noise. 
The results provide an important step for the optimization of  noise properties and thus leverage the exploitation of STNOs in prospective real applications.

\end{abstract}

\pacs{}

\maketitle

\section{Introduction}

Spin torque nano oscillators (STNOs) are nano-sized oscillators, which convert a supplying dc current into a rf electrical signal through magnetization dynamics and basic spintronic phenomena \cite{Locatelli2013}. 
Within the last decade of intensive research on spintronics, they have been identified as promising candidates for next-generation multifunctional microwave spin-electronics\cite{Locatelli2013,Ebels2017}. 
In addition to their nanometric size ($\sim 100\,$nm) and low power consumption, STNOs in general benefit from a high frequency tunability along with  compatibility with standard CMOS technology\cite{Kreissig2017-AIP,Kreissig2017-IEEE} and semiconductor manufacturing processes. 
Potential applications are manifold and go beyond the integration into future wide-band high-frequency communication systems\cite{Muduli2010,Choi2014,Purbawati2016,Ruiz-Calaforra2017,Ebels2017,Litvinenko2019}: From high data transfer rate hard disk reading\cite{Sato2012}, spin wave generation\cite{Demidov2010,Madami2011} for e.g. magnonic devices\cite{Kruglyak2010,Chumak2017}, broadband microwave energy harvesting \cite{Fang2019} or frequency detection \cite{Jenkins2016,Louis2017,Louis2018}, to  bio-inspired neuromorphic computing \cite{Torrejon2017,Romera2018}.  
Facilitating many of the mentioned potential applications, an intrinsic effect of the underlying magnetization dynamics and noteworthy the important particularity of STNOs is their non-isochronicity, i.e. the coupling between the oscillator's amplitude and phase \cite{Slavin2009}. This specifically allows tuning of the STNO frequency via the dc current. 
However, the STNO's nanoscale size and nonlinearity come at the cost of a relatively poor phase noise performance, identified as the major drawback in order to exploit STNOs in real practical applications\cite{Wittrock2019_PRB,Wittrock2020-SciRep}, especially in the field of \textit{rf} communications. 
In order to substantially improve the  phase and frequency stability, a standard means is the integration of the oscillator into  a phase locked loop (PLL) system, which on a circuit level continuously corrects the oscillator's phase fluctuations through comparison with an external reference. 
This approach applied to STNOs recently gained attention  in theory \cite{Mitrofanov2015,Mitrofanov2017} as well as  in terms of  practical realizations \cite{Keller2009,Tamaru2015,Tamaru2016,Tamaru2016-2,Tamaru2017,Kreissig2017-AIP,Kreissig2017-IEEE}. 
It represents an important step towards STNO applicability and system integration. Indeed,  PLL systems, usually implemented with a VCO (\textit{"Voltage controlled oscillator"}), are widely used in microelectronics and rf communications for various rf applications, such as clock recovery, frequency (de-)modulation, stable frequency generation, or frequency synthesis, etc. \cite{Kroupa2003,Best2007}. 
However, the design of such a system compatible to STNOs is challenging, particularly due to their strong nonlinearity, which leads to a coupling of amplitude and phase and consequently an enhancement of the phase noise. 

The PLL system used in this study has been specifically designed based  on custom integrated circuits providing a highly reconfigurable and compact system. 
Thanks to a specially designed programmable amplifier \cite{Kreissig2015} and a wide range frequency divider \cite{Kreissig2016} with high sensitivity, the PLL can be operated in a large frequency range of $0.1$-$10\,$GHz. 
In the following, we summarize the basic principle of this PLL and refer to Refs. \cite{Kreissig2015,Kreissig2016,Kreissig2017-AIP,Kreissig2017-IEEE} for further technical details. 
We demonstrate the PLL performance on two types of STNOs in different frequency ranges, i.e. magnetic tunnel junctions whose free layer adopts either the vortex state or is uniformly magnetized. 
Furthermore, a general theoretical model is developed to describe the noise characteristics of STNOs submitted to an arbitrary external signal. 
We not only consider the phase dynamics in the perfectly phase locked state, but also take phase slips into account that are a main drawback in such systems\cite{Lebrun2015,Tortarolo2018} but that have so far not yet been considered in theoretical evaluations  to correctly predict and analyze phase noise properties. 
Based on the theoretical approach, the PLL operation and  noise properties of the PLL-locked STNO are described and deterministic expressions are provided. The results from the theoretical model reproduce well the different experimentally observed features. Moreover, it provides a fundamental insight onto the complex dependencies of the PLL operation on the STNO and PLL parameters and herewith shows the routes to efficiently reduce the system's phase noise. 
Our analysis underlines the necessity to appropriately adapt the PLL to the STNO and vice versa and consequently provides the adequate parameter space to achieve that.

\section{The PLL system}
\label{sec:PLL_repetition}

\begin{figure}[bth!]
  \centering  %
  \subfloat[ \label{fig:concept_PLL-1}]
  {  %
  \includegraphics[width=0.476\textwidth]{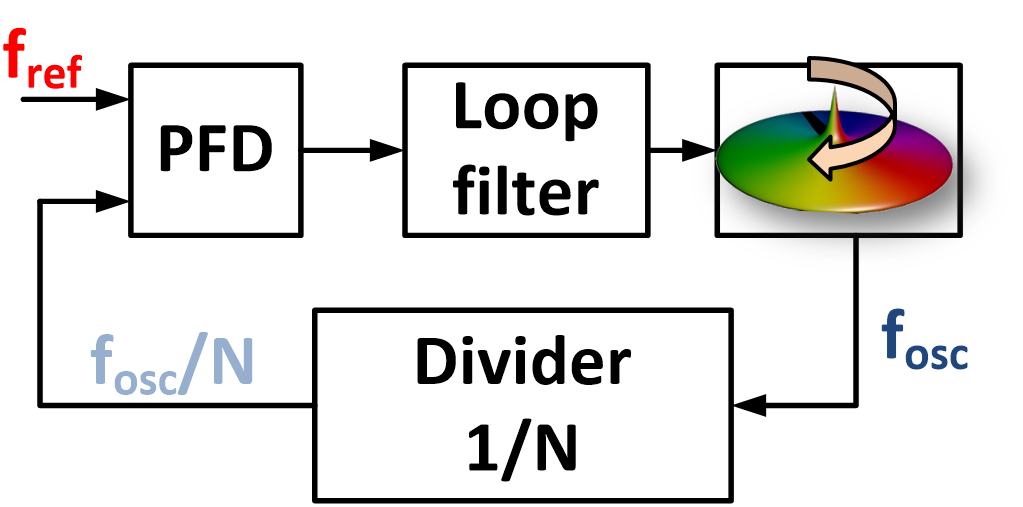}  }   
 \subfloat[ \label{fig:concept_PLL-2}] 
  { %
  \includegraphics[width=0.495\textwidth]{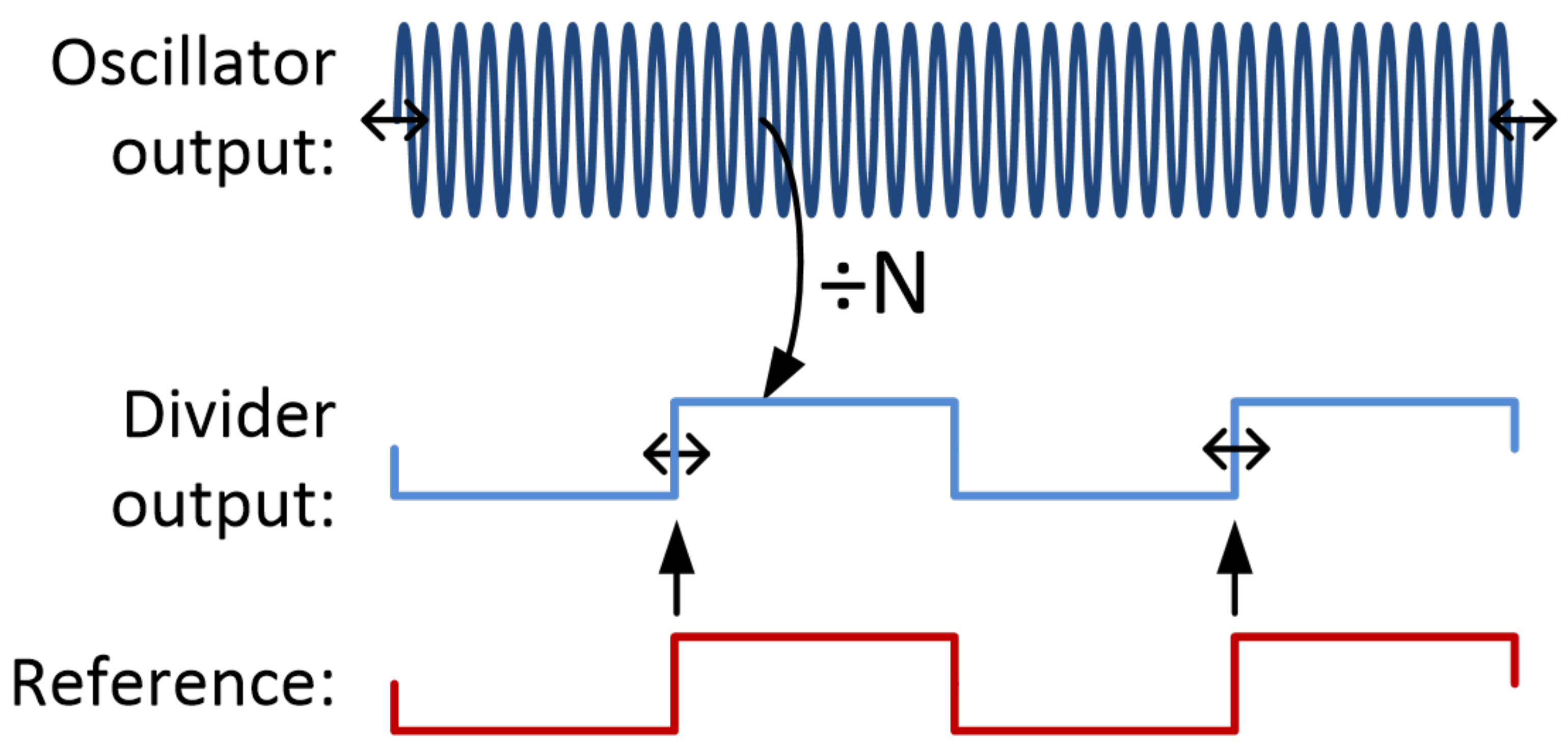}  } 
  
  \caption[Basic concept of a PLL]{Basic concept of a PLL: Stabilization of the STNO by locking the phase of the divided oscillator signal to a fixed reference. The phase difference is continuously analyzed. \textbf{(a)} Scheme of the PLL components. \textbf{(b)} Periodic signals in the PLL showing the correction mechanism.
 
}
  \label{fig:concept_PLL}
\end{figure}

The basic concept of a PLL is depicted in fig. \ref{fig:concept_PLL}. The oscillator's output signal frequency is divided by the divider ratio $N$ and subsequently compared with an external reference of frequency close to $f_{\text{ref}}\sim f_{\text{osc}}/N$. A phase frequency detector (PFD) evaluates the phase difference between these two signals whereafter the loop filter outputs a correction to the dc current $I_{dc}$  proportional to the phase difference. 
Due to its frequency tuning capability $df/dI_{dc}$, the STNO frequency is then pulled/pushed to operate at the stabilized PLL output frequency $f_{\text{osc}}=Nf_{\text{ref}}$. Its phase noise is reduced within the PLL operation bandwidth.

\begin{figure}[bth!]
\centering
\vspace{-0.0cm}
\includegraphics[width=0.6\textwidth]{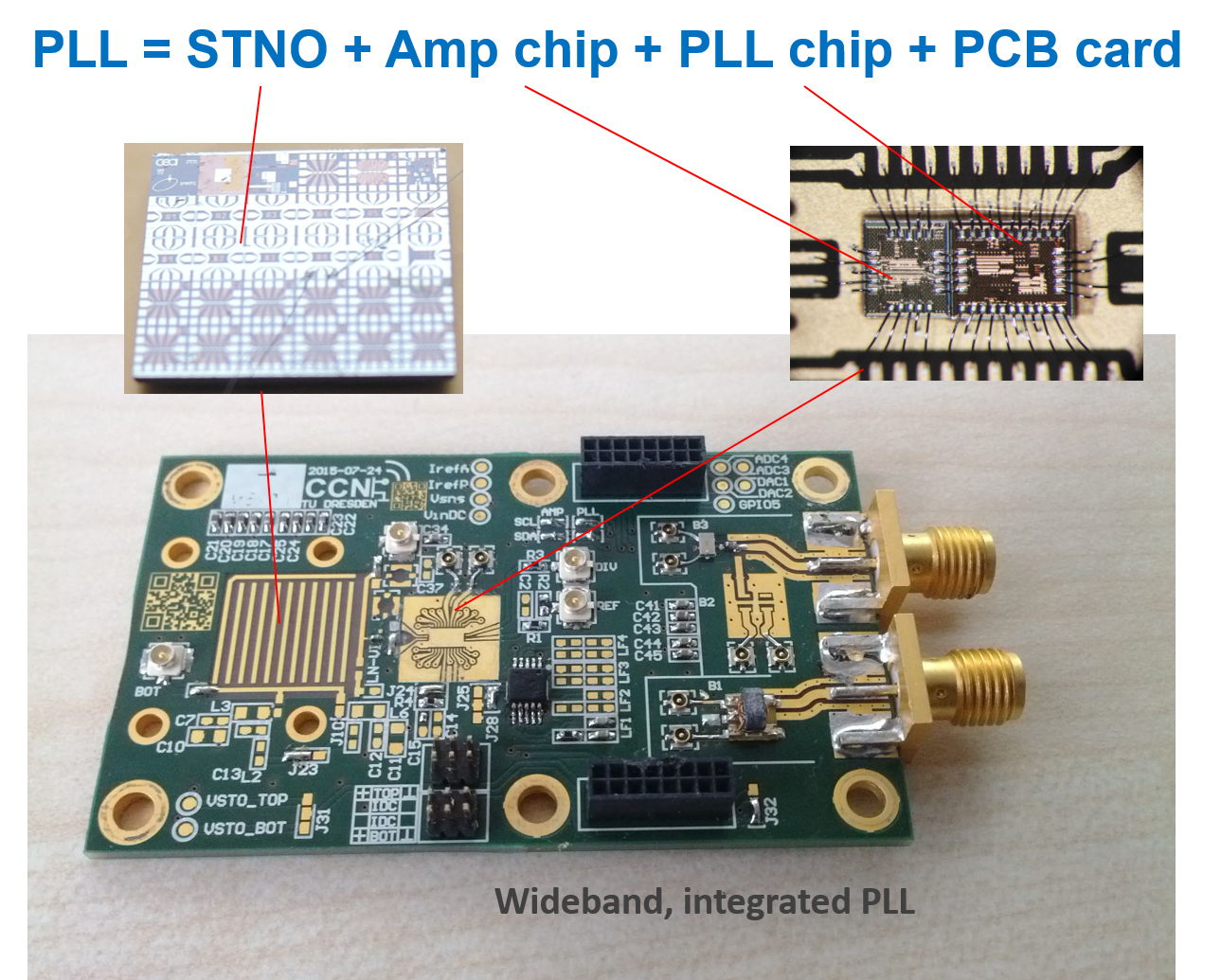}
\caption[Developed integrated PLL chip and components]{Picture of the developed PLL-chip with its components. }
\label{fig:PLL-PCB}
\end{figure}

The challenge in the circuit development of the STNO-PLL (see fig. \ref{fig:PLL-PCB})  lies in the adaptation of the  PLL concept\cite{Bellescize1932} to the specifications of STNOs. 
This particularly includes the requirement for a large PLL bandwidth, which mainly determines the ability to significantly reduce the noise characteristics and phase diffusion in the corresponding oscillator. 
Usually, PLL bandwidths in commercial rf applications range from a few kHz to a few $100\,$kHz. 
However, due to the typical noise amplitude in STNOs, bandwidths in the range of a few MHz (vortex STNOs), or up to a few tens of MHz (uniform STNOs) are necessary, although very large bandwidths often lead to detrimental spurious characteristics.  
The developed PLL system exploited in this work exhibits a bandwidth of $\sim 2.5\,$MHz. 
Furthermore, it is highly dynamic and reconfigurable. With its large range of operation frequencies of $0.1$-$10\,$GHz, it can be operated with different STNO configurations, namely in particular vortex based (STVOs) as well as uniform STNOs. 
The integrated circuit design\cite{Kreissig2015,Kreissig2016,Kreissig2017-AIP,Kreissig2017-IEEE} is shown in fig. \ref{fig:PLL-PCB}. %

\section{Experiment: PLL operation upon vortex and uniform STNOs}
\label{sec_PLL:STVO}

We present here the PLL operation for both vortex based STNOs (STVOs, operating at $300$-$500\,$MHz) and for uniformly magnetized STNOs (operating in the $5\,$GHz range). 
More experimental details are given in the appendix \ref{sec_appendix:Exp}. 

\subsection{Vortex based STNOs}

Vortex based STNOs (STVOs) rely on the spin transfer induced dynamics of a  noncollinear magnetization distribution, namely the gyrotropic motion of a magnetic vortex core \cite{Pribiag2007,Dussaux2010}. %
Typical frequencies of these devices lie in the $100\,$MHz to $1\,$GHz range.    
They exhibit large-amplitude oscillations with output powers up to the $\upmu$W range\cite{Tsunegi2014} and a good phase coherence compared to other STNO realizations. %

 \begin{figure}[hbt!]
\centering
\vspace{-0.1cm}
\includegraphics[width=0.56\textwidth]{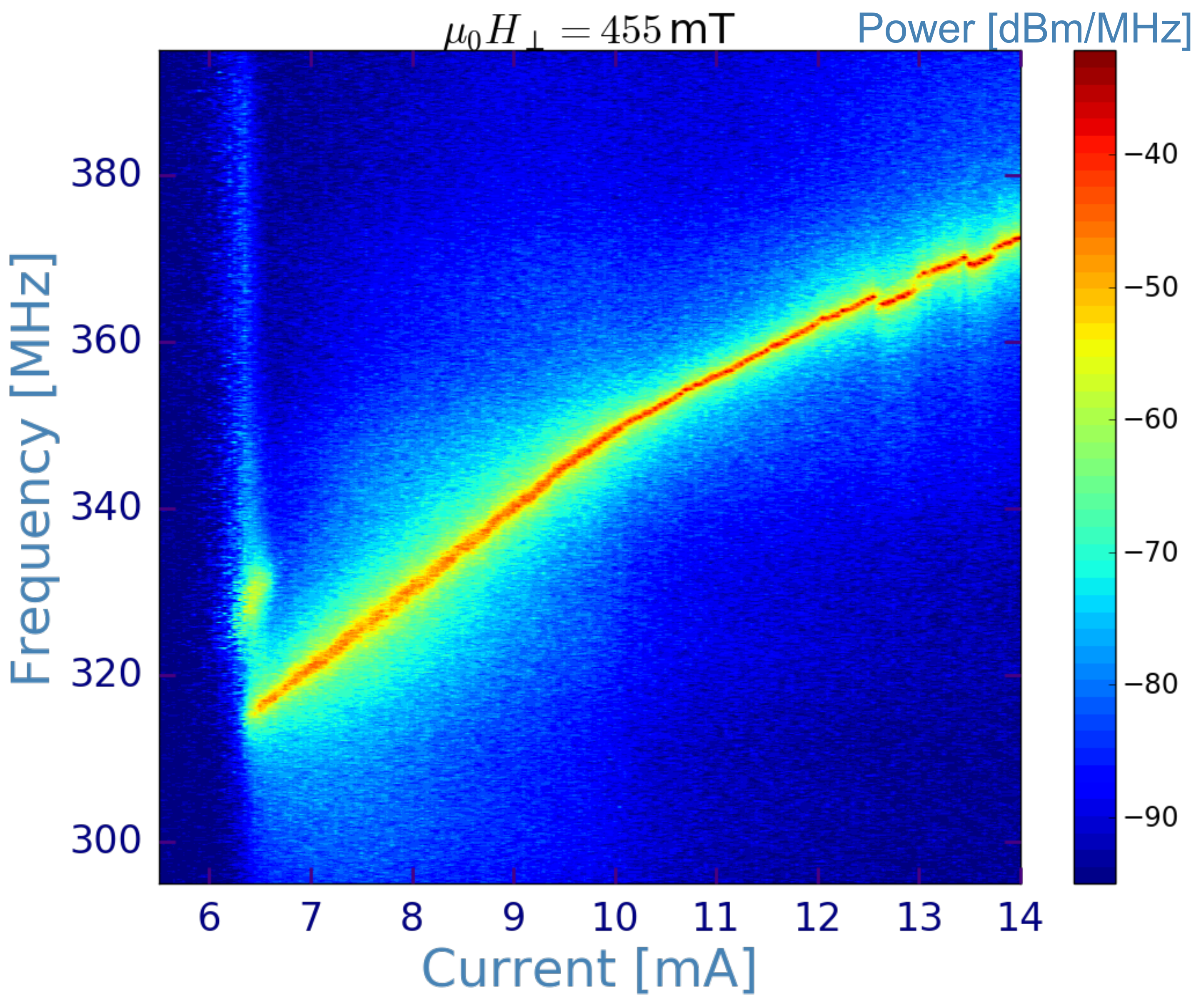}
\caption[Spectra vs. current of the measured type \textbf{b} STVO]{Frequency vs. injected current   $I_{dc}$, measured for a STVO at room temperature under a perpendicular field of $\mu_0 H_{\perp} = 455\,$mT. The color scale indicates the  emitted power. 
}
\label{fig_PLL:spectra_vs_I}
\end{figure}

An example for the  frequency characteristics of a STVO (see appendix \ref{sec_appendix:Exp} for detailed sample information) vs. the applied dc current  $I_{dc}$ is presented in fig. \ref{fig_PLL:spectra_vs_I}. 
It shows a frequency tunability of about  $df/dI_{dc}\approx 10\,$MHz/mA,  which is exploited in order to stabilize the phase noise of the sample by directly injecting  the PLL feedback current  into the STVO. %

\begin{figure}[bth!]
  \centering  %
  \subfloat[\label{fig_PLL:main-results-1}]
{%
\includegraphics[width=0.49\textwidth]{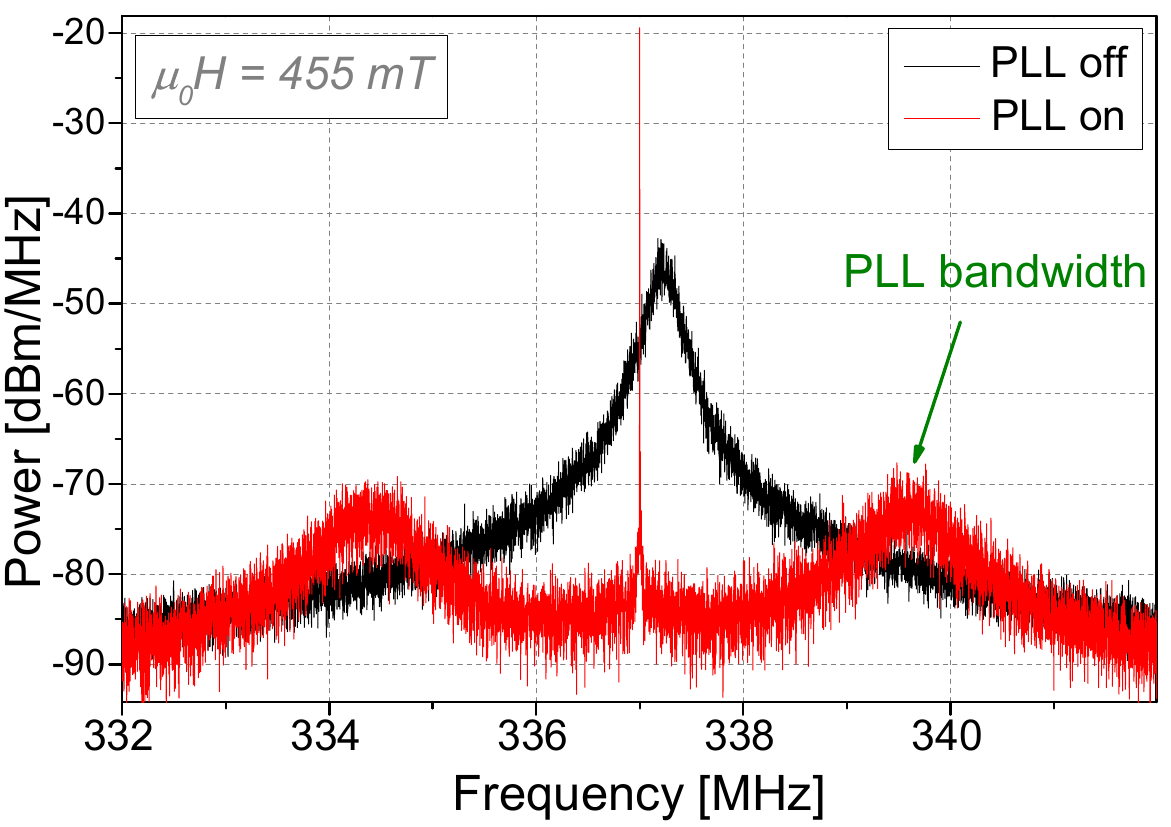}  }   
 \subfloat[ \label{fig_PLL:main-results-2}] 
  { %
  \includegraphics[width=0.49\textwidth]{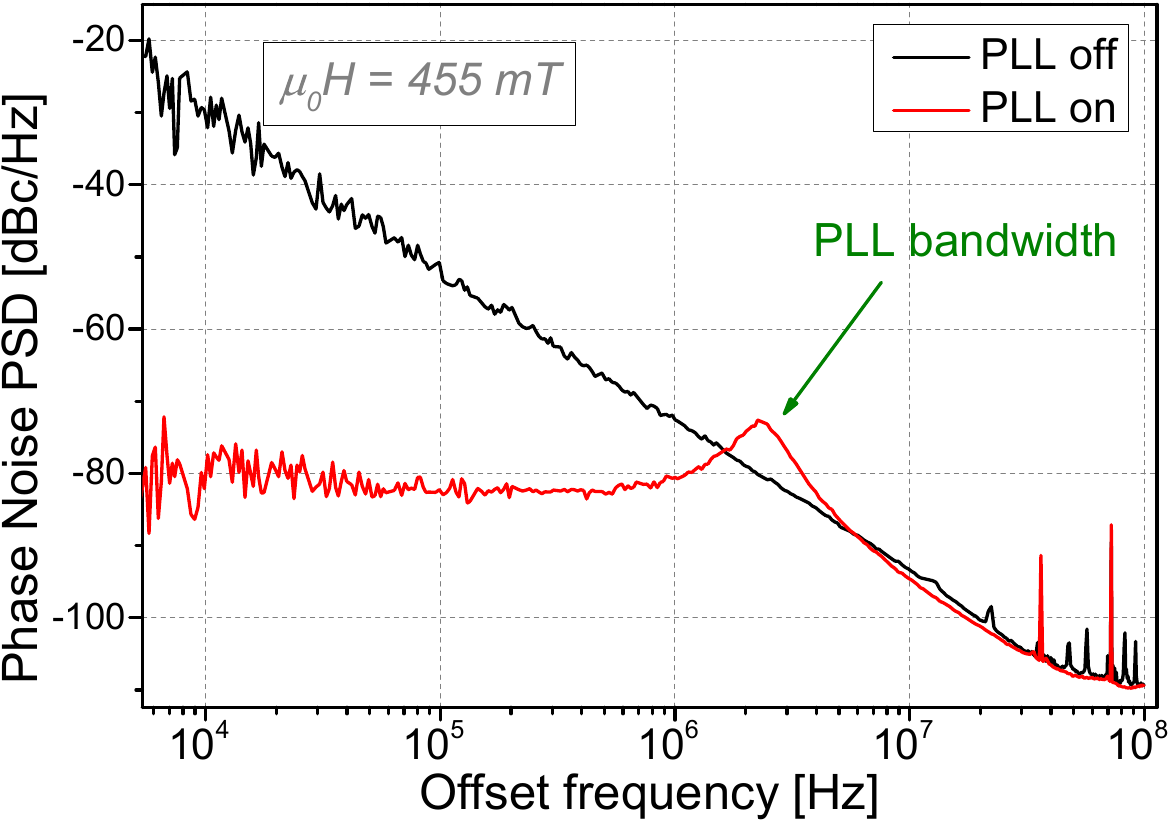}  } 
  \caption[Significant rf improvements: frequency spectra and noise PSD with/without PLL ]{ Comparison  for operation of the STVO when the PLL is off (black curve) and on (red curve) for (a) the emission radiofrequency power spectral density (PSD) and (b) the phase noise PSD. The PLL parameters are: frequency divider ratio $N=10$, charge pump $\text{CP}=8$ and coupling strength $\epsilon\sim \text{CP}/N=0.8$. Details of the experiment are provided in the appendix. 
  }
  \label{fig_PLL:main-results}
\end{figure}

In fig.  \ref{fig_PLL:main-results}, we show the emitted power spectrum  (fig. \ref{fig_PLL:main-results-1}) and phase noise\footnote{The phase noise PSD is evaluated through the Hilbert transform method, as presented more in detail in Refs. \cite{Quinsat2010,Bianchini2010,Wittrock2019_PRB}.}  (fig. \ref{fig_PLL:main-results-2}) for the
operation of the STVO device when  the PLL feedback loop  is off (black) and on (red). 
 The radiofrequency PSD peak amplitude   (fig. \ref{fig_PLL:main-results-1})  is increased by $25\,$dB up to $-19\,$dBm/MHz and  the corresponding  FWHM decreased from $\sim 1\,$MHz down to $<1\,$Hz , nominally the resolution limit of our spectrum analyzer. 
 The corresponding phase noise (fig. \ref{fig_PLL:main-results-2}) of the free running state has the typical $1/f^2$ noise dependence\cite{Grimaldi2017,Wittrock2019_PRB,Wittrock2020-SciRep} in the full offset-frequency range, while the phase noise upon PLL operation is constant within a given bandwidth $\Delta \omega_{BW}$ and hence is efficiently reduced by more than $50\,$dB at a $10\,$kHz offset from the carrier frequency.   %

\begin{figure}[bth!]
  \centering  %
  \subfloat[ \label{fig_PLL:phase-deviation-1}]
  {%
\includegraphics[width=0.49\textwidth]{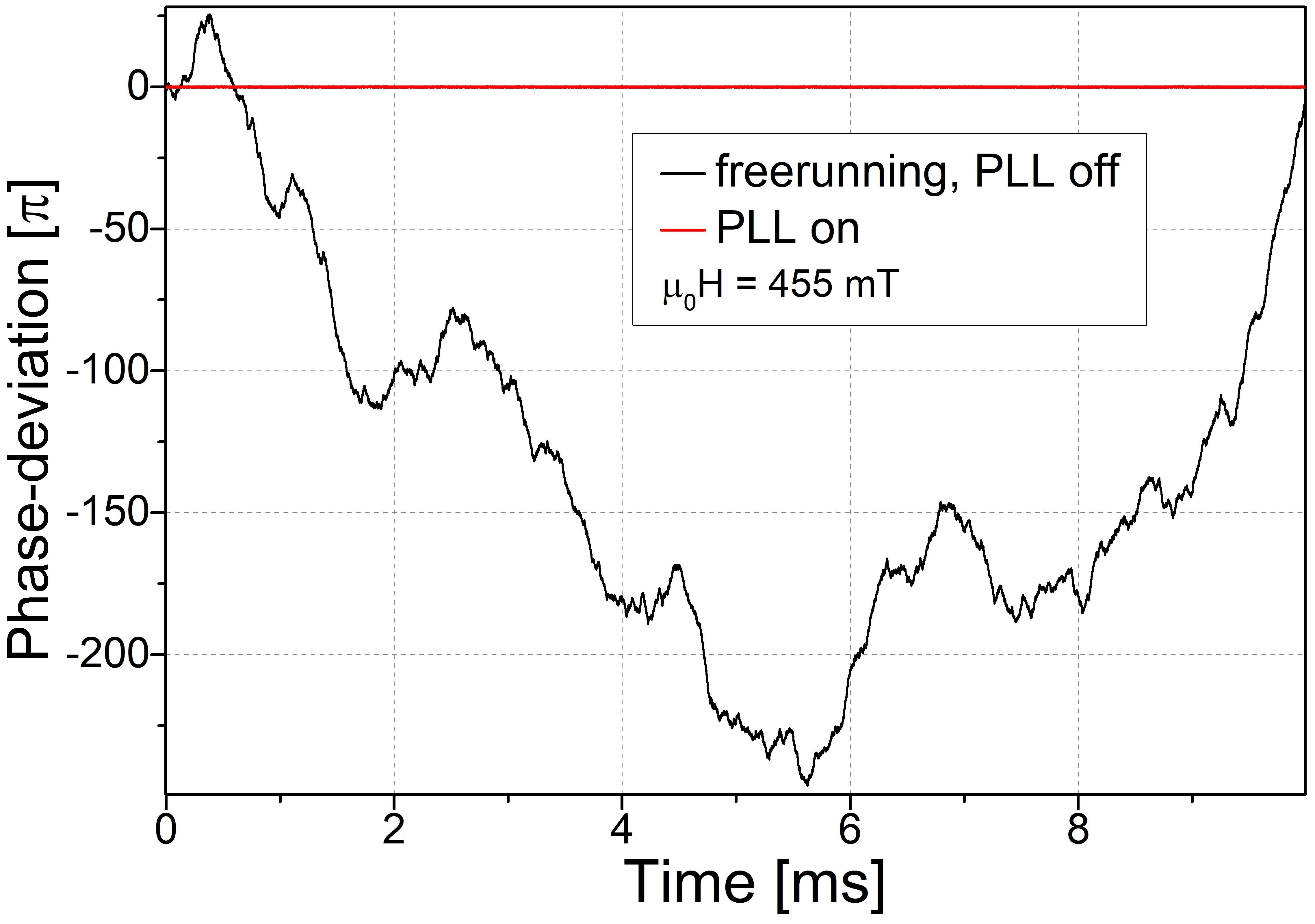}  }   
 \subfloat[ \label{fig_PLL:phase-deviation-2}] 
  { %
  \includegraphics[width=0.49\textwidth]{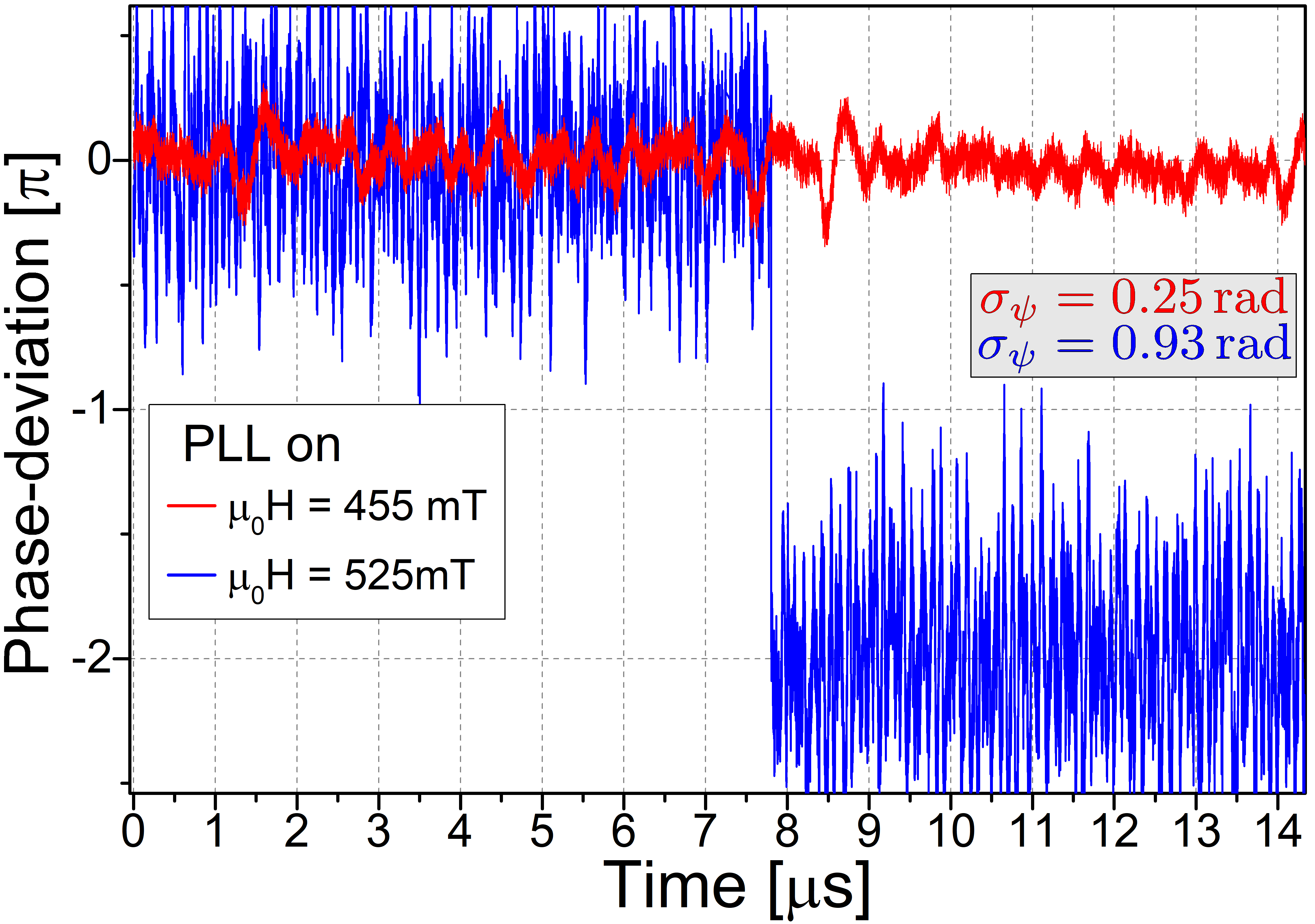}  } 
  \caption[]{Phase deviation from the mean phase.  (a) With and without PLL at $\mu_0 H_{\perp} = 455\,$mT, corresponding to the datasets shown in fig. \ref{fig_PLL:main-results}. (b) Perfect phase locking upon PLL operation (red curve, $\mu_0 H_{\perp} = 455\,$mT ) and occurrence of phase slips (blue, $\mu_0 H_{\perp} =310\,$mT). }
\label{fig_PLL:phase-deviation}
\end{figure}

The PLL operation bandwidth $\Delta \omega_{BW}$ is an important parameter, because it defines the frequency band around the carrier for which the phase deviations are detected and stabilized. 
The feedback response corrects the phase and "kicks"  it periodically on a time scale given by the inverse of $\Delta \omega_{BW}$  (see fig. \ref{fig_PLL:phase-deviation-2}). 
Hence, $\Delta \omega_{BW}$ directly correlates with the phase noise level: the higher it is, the more the phase noise can be reduced at low offset frequencies.  
 For the STNO-PLL system presented here,  it is around $\Delta \omega_{BW}/(2\pi) \approx  2.5\,$MHz (see fig. \ref{fig_PLL:main-results}) and is, as derived in section \ref{sec_PLL:theory}, related to the loop filter and the PLL parameters.

\begin{figure}[bth!]
  \centering  %
 \subfloat[ \label{fig_PLL:PLL_20_PN_conditions}] 
  { %
  \includegraphics[width=0.72\textwidth]{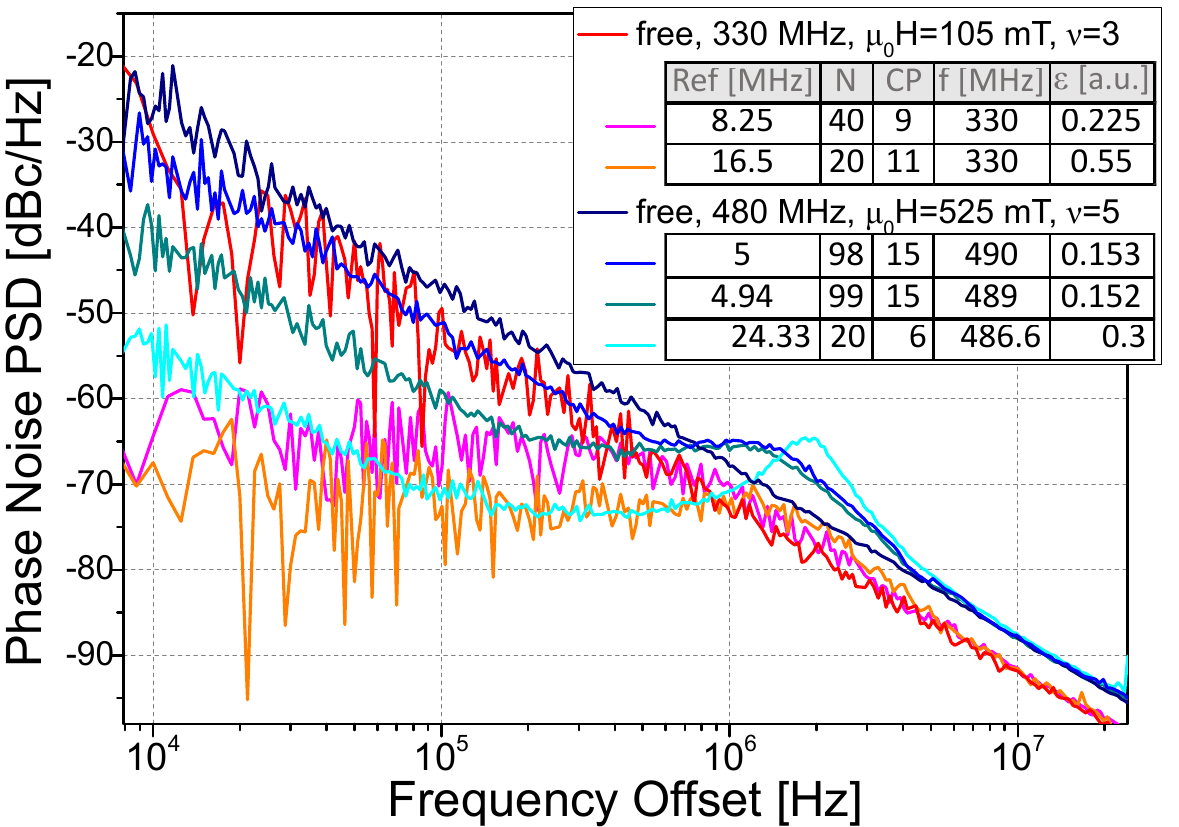}  } 
  \hfill %
 \subfloat[ \label{fig_PLL:tuning-spectra}] 
  { %
  \includegraphics[width=0.66\textwidth]{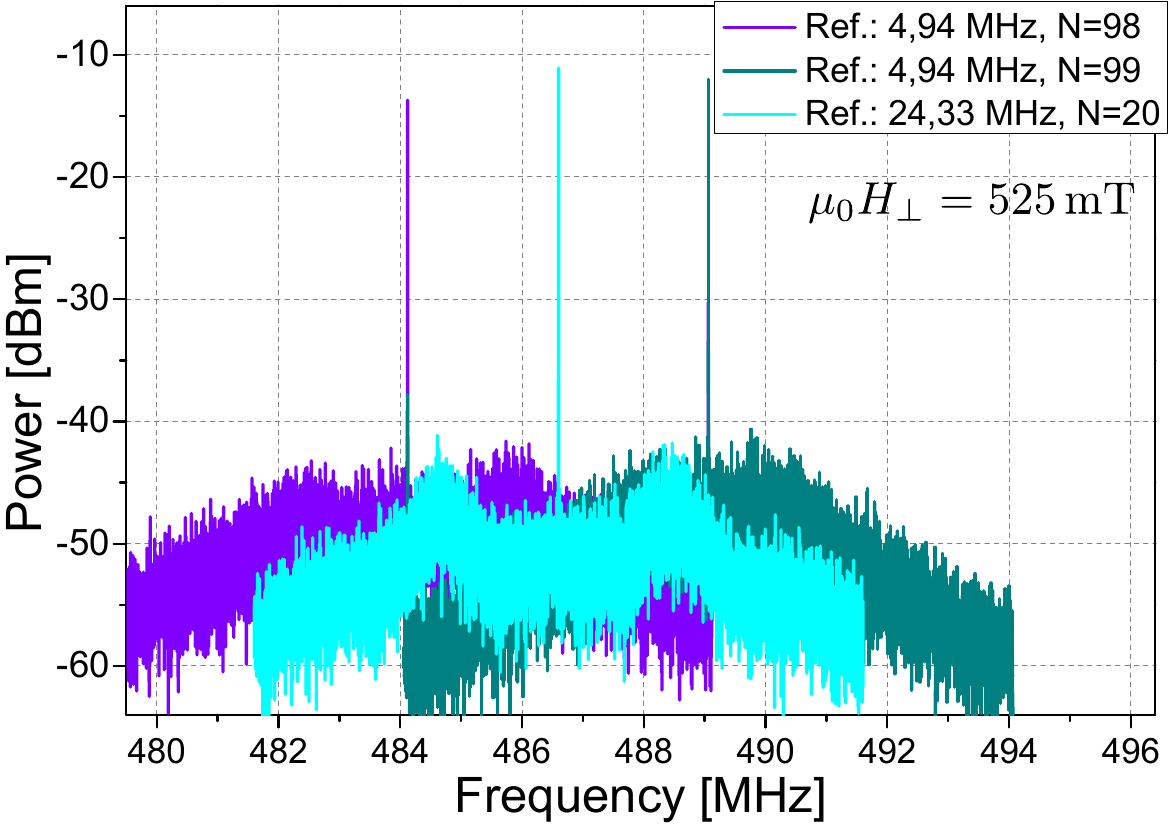}  } 
  
  \caption[Radiofrequency output characteristics of the combined system STVO + PLL for different parameter sets]{ (a) Phase noise PSDs for two free running conditions (red \& dark blue) and different parameter sets of the combined system  STVO + PLL as indicated in the inset table. (b) Power emission spectra of the locked STVO at different PLL parameters $N$ and $f_{\text{ref}}$ resulting in different central frequencies. 
Same colors in (a) and (b) match; cyan dataset with $\epsilon=0.3$ corresponds to the blue phase deviation curve in fig. \ref{fig_PLL:phase-deviation-2}. }
  \label{fig_PLL:PLL-tuning_noise_and_spectra}
\end{figure}

In fig. \ref{fig_PLL:phase-deviation}, we present the measured phase deviation from the mean phase  for the cases when the PLL is off (black) and on (red, blue) at different operation conditions. 
For the unlocked, free running STVO (black curve in fig. \ref{fig_PLL:phase-deviation-1}), the deviation to the mean phase is large and varies significantly, i.e. more than $\pm 100\pi\,$rad. 
The improvement of the STVO's coherence upon PLL operation is  highlighted by the red curve in fig. \ref{fig_PLL:phase-deviation}, for which  good  phase stabilization  within  at least $10\,$ms (total measurement time, see fig. \ref{fig_PLL:phase-deviation-1}) is achieved. 
 From  the phase deviations over time  we calculate the standard deviation $\sigma_{\psi}$ of the stabilized phase for the phase locked state. 
For PLL on and an applied field of $455\,$mT, we find $\sigma_{\psi}=0.25\,$rad which is in good agreement with values obtained using commercial PLLs,  lying between $\sigma_{\psi}=0.237\,$rad and $\sigma_{\psi}=0.923\,$rad \cite{Tamaru2017,Tamaru2016,Tamaru2016-2}.
However, at some operation conditions, the PLL does not achieve a complete phase locking but instead, phase slips by $2\pi$ at some instances in time occur in the phase deviation, as shown by the blue curve in fig. \ref{fig_PLL:phase-deviation-2}. 
They occur when the frequency divider raises  a counting error due to too large fluctuations above $\Delta \omega_{BW}$ (see Ref. \cite{Tamaru2016}).  
The phase slips  reflect a diffusive phase noise process and hence, manifest themselves as a $1/f^2$ contribution in the low offset frequency range of the phase noise PSD, as observed for  instance in fig. \ref{fig_PLL:PLL_20_PN_conditions} for the blue-colored curves  (note, the cyan-colored curve is based on the same measurement data as the blue curve in fig. \ref{fig_PLL:phase-deviation-2}).

In order to better assess  how the STNO and PLL parameters and the phase jumps affect the PLL performance,  we present in fig. \ref{fig_PLL:PLL_20_PN_conditions}, a series of phase noise (PN) PSDs for different PLL parameter settings, such as the frequency divider ratio $N$, reference signal frequency $f_{\text{ref}}$ and coupling strength $\epsilon\sim \text{CP}/N$, with CP the charge pump of the PLL feedback as a product of correction current and driver gain\cite{Kreissig2017-IEEE}. 
It can clearly be seen that the locking and desynchronization of the STNO inside the PLL circuit strongly depend on the PLL parameter settings along with the parameters that determine the free running STNO performance (dc current and applied magnetic field). 
In fig.  \ref{fig_PLL:PLL_20_PN_conditions}, phase noise PSDs are shown for zero frequency mismatch between free running ($f=330\,$MHz) and PLL frequency and for the two cases when the  PLL is turned off (red) as well as for the PLL turned on (pink, orange). 
For this  set of PLL parameters, the STVO is well locked to the PLL in the frequency range below the bandwidth $\Delta f_{BW} = \Delta \omega_{BW}/(2\pi)$, where the phase noise is flat and well reduced. 
Furthermore, it is observed that upon increasing of the PLL coupling $\epsilon \sim \text{CP}/N$ the phase noise is more efficiently decreased. \\
For another set of data (blue-colored  curves in fig. \ref{fig_PLL:PLL_20_PN_conditions}) with non-zero frequency mismatch, we find that phase slips become important. For instance for the  cyan curve with $N=20$ ($f_{\text{ref}}=24.33\,$MHz), $\epsilon=0.3$ and a frequency mismatch of $6.6\,$MHz, a significant $1/f^2$ noise contribution appears for offset frequencies below the corner frequency (transition between $1/f^0$ and $1/f^2$ curve shapes) $f_{\text{corner}} \approx 10^5\,$Hz,  while the phase noise is efficiently reduced in the frequency range between the PLL bandwidth $\Delta f_{BW}$ and $f_{\text{corner}}$.  
For lower coupling $\epsilon = 0.153$, $\epsilon=0.152$ %
combined with a larger frequency mismatch  of $10\,$MHz and $9\,$MHz respectively,  the phase noise reduction is less efficient, i.e. phase jumps become more frequent. 
In consequence, the $1/f^2$ noise levels are higher and the corner frequency   $f_{\text{corner}}$ approaches the PLL bandwidth $\Delta f_{BW}$. 
Furthermore, comparing the two curves of similar coupling strength $\epsilon = 0.153$ (blue) and $\epsilon=0.152$ (green-blue), the blue one with larger frequency mismatch shows a larger phase noise level. 
This demonstrates also the importance of the frequency mismatch for the occurrence rate of the phase slips. 
A larger nonlinear amplitude-phase coupling, quantified by the parameter $\nu$ (see $\nu=5$ vs. $\nu=3$ in fig. \ref{fig_PLL:PLL_20_PN_conditions}), favours the occurrence of phase slips and also leads to a generally larger noise level at higher frequency offsets. 
Comparing the datasets in fig. \ref{fig_PLL:PLL_20_PN_conditions} for zero and non-zero frequency mismatch, one more important feature needs to be pointed out. For the latter set (blue-colored curves), a small resonance peak is visible around the bandwidth frequency. Fig. \ref{fig_PLL:PLL_20_PN_conditions}  suggests that its amplitude, and as well the bandwidth itself, scale with the PLL coupling parameter $\epsilon$ and the non-linear coupling parameter $\nu$. \\
 As a final point, we demonstrate in fig. \ref{fig_PLL:tuning-spectra}, that the developed PLL is very versatile and can also be used for frequency synthesis. Notably, its  output frequency can be easily shifted by several MHz through either the PLL divider ratio $N$ and/or the reference frequency $f_{\text{ref}}$.

To conclude, the PLL measurements on vortex STNOs reveal several characteristic features, such as the PLL bandwidth $\Delta \omega_{BW}$, the phase noise reduction in the locked state, the phase slips leading to a $1/f^2$ contribution and a resonance around $\Delta \omega_{BW}$. 
They depend on the free running STNO performance and parameters as well as on the PLL settings. These complex dependencies will be  further discussed in sections \ref{sec_PLL:theory}-\ref{sec:discussion_PLL} where theoretical expressions for the phase noise of the PLL-STNO system are derived considering stochastic white noise.

\subsection{Uniform STNOs}
\label{sec_PLL:uniform}

To demonstrate the developed PLL chip's large operational frequency range of $0.1$–$10\,$GHz, we also operate the PLL with other available STNO devices. 
These are in-plane quasi-uniformly magnetized magnetic tunnel junctions (MTJs),  composed of an in-plane magnetized reference layer and an in-plane magnetized free layer. The magnetic stack is very close to the vortex MTJs but with a much thinner free layer to stabilize the uniform in-plane magnetization. The  applied in-plane magnetic field  is adjusted for operation around $4$-$5\,$GHz, leading to  a free running emission power of $10$-$50\,$nW and a linewidth of $10$-$20\,$MHz.

\begin{figure}[bth!]
  \centering  %
  \includegraphics[width=0.764\textwidth]{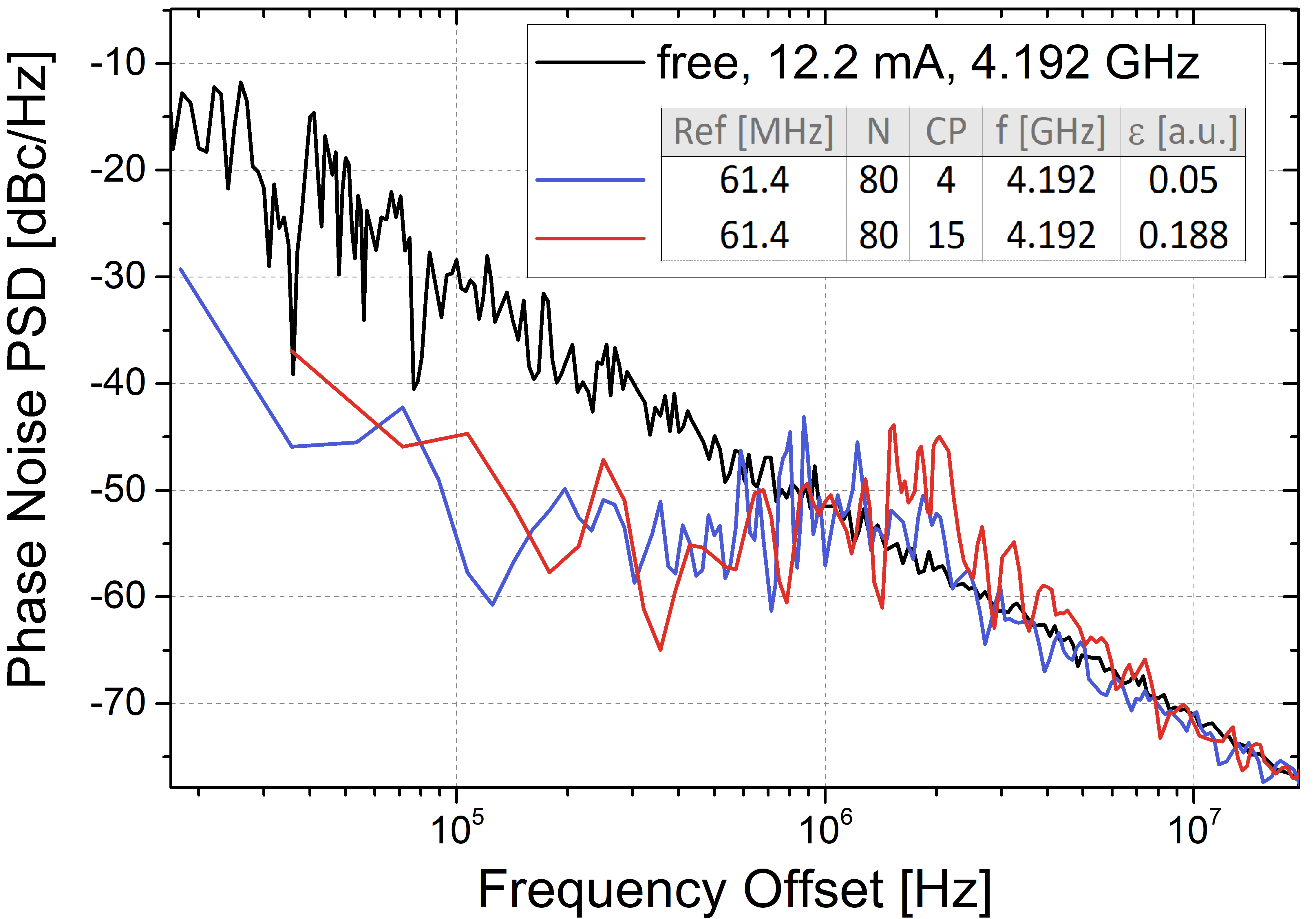} %
   
  \caption[]{Measurement of a uniform STNO in the PLL circuit: Phase noise spectral density for the free running and the locked oscillator at different PLL parameters. } %
  \label{fig_PLL:uniform-noise-1}
\end{figure}

Time domain characterization   of  the free running output voltage signal of this  in-plane (IP) STNO  reveals that the oscillations are not stable in time, with extinctions (strong amplitude reductions and possible loss of phase coherence) over short time scales ($1\,$ns). The average time of stable oscillations was $\sim 0.1$ - $1\,\upmu$s.  Large amount of extinctions result in an increased phase noise of the free running state as compared to the vortex  and other uniform STNO devices\cite{Tamaru2016,Tamaru2016-2}.  Furthermore, the free running linewidth is larger than the PLL bandwidth. This altogether makes the operation of the PLL with the available uniform magnetized STNOs more difficult and lowers the locking capability. Notably we find that the PLL was locked only over finite time scales, in the best cases this was $20$-$60\,\upmu$s. Therefore, when evaluating the phase noise over the full measured time trace for devices with very high number of locking failures (i.e. locking times much below $100\,\upmu$s), we observe a phase noise reduction of only $12\,$dB at $100\,$kHz offset frequency for a PLL bandwidth of $2\,$MHz. 
However, when evaluating the phase noise for the case of low number of locking failures and locking times of $20$-$60\,\upmu$s (see fig. \ref{fig_PLL:uniform-noise-1}), the phase noise reduction at $100\,$kHz offset frequency increases to $20\,$dB, the noise plateau is well pronounced and the resonance around the PLL bandwidth frequency is well visible at large coupling. 
Nevertheless, when increasing  the PLL coupling $\epsilon = \text{CP}/N$ in this case, similar features can be observed as for the PLL operation with a STVO: occurrence of a plateau, corner frequency below which the $1/f^2$ phase diffusion sets in and a more pronounced ''bump'' around the PLL bandwidth. 
To conclude, the phase noise of the vortex and uniform STNOs reveals very similar features in the PLL phase-locked state and the designed PLL system is  operational for both device configurations, i.e. for a large frequency range.

\section{Theory: PLL phase dynamics including stochastic white noise and phase slips}
\label{sec_PLL:theory}

In addition to the experimental results, we develop here a complete theoretical approach in order to assess the noise properties of STNOs, especially when they are subject to an external signal. 
This includes, i.a., the STNO synchronization to fractional or harmonic frequencies, frequency modulation, or finally the exploitation of a PLL onto the STNO. 
We do not restrict to the assumption of perfect phase locking but, importantly, also take the occurence of phase slips into account, which have not yet been treated in this context. 
However, as we present experimentally in sec. \ref{sec_PLL:STVO}, they are the main drawback in the synchronization of a STNO. 
With our theoretical approach, we are able to give deterministic expressions that fully determine the characteristics and noise properties of STNOs upon an external forcing. 
In section \ref{sec:discussion_PLL}, we discuss the results with a focus on the operation of a PLL but stress, that the discussion remains general and valid also for other forcing signals. 
A comparison  between the experimental results of section \ref{sec_PLL:STVO} and the developed theoretical expressions allows us to fully explore the complex parameter space of the PLL (but not limited to that) upon a STNO in detail.

We first present the phase equations for non-isochronous auto-oscillators such as STNOs, including external forcing and noise. 
From this, we derive an expression for the noise power spectral density (PSD)  for  perfect phase locking. Subsequently, we consider the occurrence of phase slips based on the Brownian motion inside the synchronization potential of the locked STNO.

\subsection{Power and phase equations for the auto-oscillator under external forcing}

The nonlinear auto-oscillator theory\cite{Tiberkevich2007,Slavin2009} is used to describe the STNO phase equations under external forcing.  
It describes the oscillations through the complex  amplitude $c(t)= \sqrt{p(t)} e^{-i\phi(t)}$, with $p(t)=|c|^2$ being the oscillation power and $\phi(t)$ the phase: 
\begin{align}
\dot{c} + i \omega (|c|^2) c + \Gamma_{\text{eff}}(|c|^2) c = f(t) ~~~.   \label{eq:Slavin}
\end{align}
The oscillation angular frequency is denoted by $\omega$ and $\Gamma_{\text{eff}} = \Gamma_+ - \Gamma_-$ is the effective damping rate with positive rate $\Gamma_+$ representing the losses of the system, and negative one $\Gamma_-$  representing the system's gain. 
 $f(t)$ is a  function, which allows a description of the system's interaction with the environment, e.g. including noise processes and an injected external signal. 
 Due to the dependence of the damping terms on the amplitude ($\frac{d\Gamma_-(p)}{d p} < 0$ and $\frac{d\Gamma_+(p)}{d p} > 0$), the oscillation is described by a limit cycle with stable oscillation power $p_0$, which is obtained when the positive and negative damping terms equal: $\Gamma_-(p_0)=\Gamma_+(p_0)$. 
Assuming a small perturbation $\delta p$ of the stable oscillation power $p(t)=p_0+\delta p$ due to noise yields a characteristic damping rate $\Gamma_p = \pi f_p = \left[ \frac{d\Gamma_+}{dp} (p_0) - \frac{d\Gamma_-}{dp} (p_0) \right] p_0$ of small power deviations back to the stable limit cycle\cite{Tiberkevich2007,Slavin2009}. 
The parameter $\nu = d\omega(p)/dp \cdot p_0/(\pi f_p)$ %
is the normalized dimensionless nonlinear frequency shift and quantifies the coupling between phase and amplitude due to nonlinearity.

In order to study the system for an injected external signal, we add a periodic interaction term $f(t) = \Omega a_{\text{ext}} e^{-i \left( \beta +  \phi_{\text{ext}} \right) }$ to eq. (\ref{eq:Slavin}). 
  At first, the external source is assumed spectrally pure and noise-free; $\Omega $ %
  denotes the coupling strength to the oscillator, $a_{\text{ext}}$ the injected signal amplitude, $\beta$ the coupling phase, which might be different for amplitude and phase ($\beta_p$ and $\beta_{\phi}$, resp.) and which can also account for a time delay, and $\phi_{\text{ext}}$ the phase of the external signal. 
  Explicitly writing eq. (\ref{eq:Slavin}) for the phase and power separately\cite{Wittrock2019_PRB,Slavin2009}, the system can be linearized around $p_0$.  Upon  renormalization $\rho = (p-p_0)/(2p_0)$ of small deviations from $p_0$ due to the external signal,  one obtains:
\begin{subequations} 
 \begin{align}
 \dot{\rho}  &=  -2 \Gamma_p \rho +  \epsilon_p  a_{\text{ext}} \cos( n \psi - \beta_p  )  \\
 \dot{\psi} &=  \Delta\omega - {\epsilon}_{\phi}  a_{\text{ext}} \sin( n \psi - {\beta}_{\phi} ) + 2 \nu \Gamma_p {\rho}   ~~. 
 \end{align} \label{eq_theo:mutual_1}
 \end{subequations}
We introduce here the phase difference $\psi =  \phi - \phi_{\text{ext}}/n $ and frequency mismatch $\Delta \omega = \omega - \omega_{\text{ext}}/n$ between STNO and external signal and the normalized coupling  $\epsilon_{p,\phi} =     {\Omega_{p,\phi} }/{\sqrt{p_0}}$. 
We assume the nonlinear damping rate larger than the coupling $\Gamma_p \gg \epsilon_{\phi}$. The parameter $n \in \mathbb{Q}$ accounts for either a possible fractional synchronization\cite{Urazhdin2010,Li2011,Lebrun2015} or  for the frequency division $n= 1/N$ of the PLL system. 

From the stationary solutions to eqs. (\ref{eq_theo:mutual_1}), the equilibrium phase difference $\psi_{eq}$ and power deviation $\rho_{eq}$ can be derived. 
For the equilibrium state to be stable, all the eigenvalues of the Jacobian $J$ of the dynamical system (\ref{eq_theo:mutual_1}) must be negative. The latter is given by:
\begin{align}
J = \begin{pmatrix}
-2 \Gamma_p  &  -n \epsilon_p a_{\text{ext}} \sin(\beta_p + n \psi_{eq})  \\
 2\nu \Gamma_p  &  -n \epsilon_{\phi} a_{\text{ext}} \cos(\beta_{\phi} + n \psi_{eq})
\end{pmatrix}  ~~~.    \label{eq_PLL:Jacobian}
\end{align}
\hspace{0.25cm} and fundamentally characterizes the system's stability. 
Note that the theoretical approach in this section is rather general and indeed valid for any external source applied to the STNO.

\subsection{ General form of the noise PSD at efficient phase locking and white noise}

As elaborated in the experiments (sec. \ref{sec_PLL:STVO}), the forced oscillator can respond in two ways to stochastic noise processes. 
First, when it is locked to a stable phase difference $\psi_{\text{eq}}$, the fluctuations occur inside a synchronization potential valley around this value. 
The second mechanism describes discrete phase jumps between two $2\pi/n$ periodic equilibrium states, separated through a potential barrier.  
In this section, we first consider only the locked case while in section \ref{sec_PLL:phase-slips}, also the occurrence of phase slips will be taken into account.

In the presence of thermal noise, $\psi$ and $\rho$ are subject to small variations around their stationary values $\psi_{\text{eq}}$ and $\rho_{\text{eq}}$. 
The variation dynamics can be described in linear algebra and response theory by the time-independent matrices $A=-J$ and $\Sigma = \sigma_0 \mathbbm{1} $ through a linear stochastic differential equation with additive noise\cite{Risken1989}: 
\begin{align}
\dot{X}_t = -A X_t + \Sigma H_t ~~~.  \label{eq_PLL:matrix-SDE}
\end{align}
$X_t =  \left( \delta \rho , \delta \psi \right)^{\intercal}$ is a Gaussian stationary Ornstein-Uhlenbeck process, $\Sigma$ describes the correlation matrix of the normalized amplitude and phase noise processes $H_t = \left( \eta_p , \eta_{\phi} \right)^{\intercal}$ with diffusion constant $\sigma_0$ (with the diffusion coefficient $D_n$ of the Gaussian noise process, it is $\sigma_0^2 = 2 D_n(p_0)/p_0 = 2 \Delta\omega_{0}$ with $\Delta\omega_{0}$ the STNOs linear half-linewidth\cite{Tiberkevich2007,Kim2008}).
By evaluating the eigenvalues of $A$, a classical stability analysis can be performed. 
Transformation into the frequency space and exploiting the characteristics of the Gaussian noise process yields the phase noise PSD (for details see appendix \ref{sec_appendix:PSD-calculus}):

\begin{align}
S_{\delta \psi^2} (\omega) = \frac{\sigma_0^2}{4\Gamma_p^2}  \frac{1+ \nu^2 + v^2}{ \left( \tilde{\lambda}_1^2 + v^2  \right)  \left( \tilde{\lambda}_2^2 + v^2  \right)}   ~~~,    \label{eq_PLL:PN}
\end{align}
\hspace{0.25cm} with $\tilde{\lambda}_{1,2}=\lambda_{1,2}/(2\Gamma_p)$ and $v = \omega/(2\Gamma_p)$ normalized quantities  and the eigenvalues of the system matrix $A$ given by: 
\begin{align*}
\lambda_{1,2} = \Gamma_p \left( 1+u \right)  \left( 1  \pm  \sqrt{ 1 -  4 \gamma  }   \right)   ~~~.
\end{align*}
We have defined the following variables: 
\begin{align}
u =& \frac{n{\epsilon}_{\phi} a_{\text{ext}}}{2\Gamma_p} \cos(  n \psi_{eq} - {\beta}_{\phi} )  \nonumber  \\
\gamma =& \frac{\det(A)}{\left(\text{tr}(A)\right)^2} =  \frac{a_{\text{ext}} n}{2 \Gamma_p (1+u)^2}  \nonumber \\
& \cdot \left(  {\epsilon}_{\phi} \cos(  n\psi_{eq} - {\beta}_{\phi} )  + \nu   \epsilon_p \sin(  n\psi_{eq} - \beta_p )            \right)  ~~~ . \label{eq:gamma_u}
\end{align}

It can be seen that $\gamma$ determines the eigenvalues to be real or complex valued. 
A discussion of the result \eqref{eq_PLL:PN} is given in section \ref{sec:discussion_PLL}.

 \subsection{Phase slip dynamics}
\label{sec_PLL:phase-slips}

\begin{figure}[hbt!]
\centering
\vspace{-0ex} %
\includegraphics[width=0.52\textwidth]{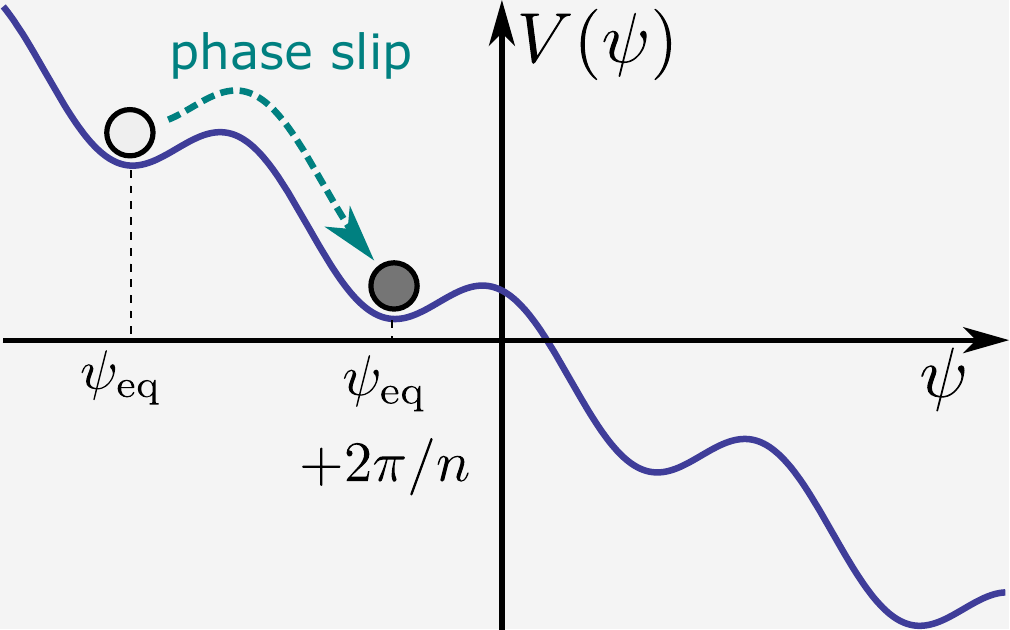}
\caption[Synchronization phase slip dynamics in a periodic potential]{Representation of the  phase slip dynamics in the periodic synchronization potential $V(\psi)$. } %
\label{fig_PLL:potential-field-1}
\end{figure}

As seen in  section \ref{sec_PLL:STVO}, phase slips by $\pm 2\pi/n$ are one of the major drawbacks in terms of STNO performance in general and of STNO performance within a PLL system in particular. They add an additional diffusive contribution to the noise PSD. 
Hence in the following, we study the physics of these phase jumps similar to a Brownian motion in a periodic  potential field $V(\psi)$ \cite{Stratonovich1965,Reimann2002}. %
Within such a model, the  dynamics of the phase difference $\psi$ can be characterized by an effective drift velocity $v_e$ and diffusion constant $D_e$, defined asymptotically through\cite{Stratonovich1965,Reimann2002}:
\begin{align}
v_e = \lim_{t\rightarrow \infty} \frac{\braket{\psi(t)}}{t} ~;  \quad  D_e = \lim_{t\rightarrow \infty} \frac{\braket{\psi^2(t)} - \braket{\psi(t)}^2}{2t}  ~.  \label{eq_PLL:drift-diffusion_time-limes}
\end{align}
In terms of phase noise, this diffusive mechanism translates to the described $1/f^2$ shaped noise characteristics:
\begin{align}
    S_{\delta\psi^2}(\omega)= \frac{D_e}{\omega^2} \label{eq:noise-PSD_phase-slips} ~~~.
\end{align}
The phase slip mechanism within the synchronization potential $V(\psi)$ is shown in fig. \ref{fig_PLL:potential-field-1}, where thermal fluctuations induce a phase slip to another minimum. 
The drift and diffusion due to the phase slips are determined by the potential barrier between two minima in $V(\psi)$, the frequency mismatch $\Delta \omega$ between STNO and external signal, whose sign sets the favored drift direction, and  the intrinsic diffusion $D$ of the phase difference.  
 We derive expressions for these parameters and describe the fluctuation dynamics within the Fokker-Planck formalism \cite{Moss1989,Risken1989}.

Under the assumption that phase slips occur at time scales larger than the power relaxation rate $\tau\gg 1/\Gamma_p$, 
the derivative $\dot{\rho}$ of the amplitude deviation  in the coupled eqs. (\ref{eq_theo:mutual_1}) is small compared to  $\Gamma_p \rho$ and hence $\rho$ can be expressed as $\rho (\psi (t))$. 
Inserting this into the stochastic differential equation for the dephasing $\psi$ (eqs. (\ref{eq_theo:mutual_1})  including noise), one obtains:
\begin{align}
\dot{\psi} =& \Delta\omega - \epsilon_{\phi} a_{\text{ext}}  \sin( n\psi - \beta_{\phi}) + \nu \epsilon_p a_{\text{ext}} \cos( n \psi - \beta_p ) \nonumber \\[1ex]
 & + \underbrace{\sigma_0 \eta_{\phi} + \sigma_0 \nu \eta_{p} }_{\sigma_0 \sqrt{1+\nu^2} \eta }   ~~~.   \label{eq_PLL:Adler_for_PLL}
\end{align}
The term under the curly bracket renormalizes the noise terms and defines the diffusion $D=(1+\nu^2) \sigma_0^2/2 $ of the phase dynamics. Eq. (\ref{eq_PLL:Adler_for_PLL}) corresponds to the stochastic Adler equation\cite{Pikovsky2003-synchronization,Balanov2008-synchronization} %
and  defines the periodic synchronization potential $V(\psi)$ (see appendix \ref{sec_appendix:phase-slip-dynamics}) with local minima at  $\psi = \psi_{eq} +  2\pi k/n$, $k\in\mathbb{Z}$.

In the Fokker-Planck formalism\cite{Moss1989,Risken1989} upon the potential $V(\psi)$, the  $2\pi/n$ periodic  probability density $\mathscr{P}(\psi , t)$  of the dephasing value $\psi$ at time $t$ describes the fluctuation dynamics (see appendix \ref{sec_appendix:phase-slip-dynamics} for details). 
Analytical expressions for the parameters $v_e$ and $D_e$ can then be found from the stationary probability density $\mathscr{P}_0 (\psi)$,  evaluated through the corresponding continuity equation with  probability current $j_{\mathscr{P}}$ (see appendix \ref{sec_appendix:phase-slip-dynamics} for details). 
They can be given as:
\begin{align}
v_e = \frac{nD}{\pi} \sinh\left( \frac{\pi F}{2} \right) \left| I_{iF/2} \left( E/2 \right)   \right|^{-2}   ~~~,  \label{eq_PLL:phase-drift_equation} \\
\intertext{and}
D_e = D \cosh\left(  \pi F/2       \right)  \left| I_{iF/2} \left( E/2 \right)   \right|^{-2}  ~~~. \label{eq:diffusion_coeff}
\end{align}

\hspace{0.25cm} with $I_{ia}(x)$ the modified Bessel function \cite{Bronstein}, $(a,x)\in \mathbb{R}$, $E = \left[ 2a_{\text{ext}} (\nu \epsilon_p + \epsilon_{\phi})\right]/(nD)$ a normalized coupling, and the normalized frequency mismatch $F=2\Delta\omega/(nD)$. 
We treat the diffusion constant $D_e$ using the approach presented by  Stratonovich\cite{Stratonovich1965} assuming a small drift (frequency detuning $\Delta \omega$ much smaller than the periodic coupling terms in (\ref{eq_PLL:Adler_for_PLL})). 
Note that in general, the evaluation of $D_e$ is rather complex. 
However, the derived relation is a good approximation for small normalized frequency detuning $F=2 \Delta\omega/(nD) \ll 1$ \cite{Stratonovich1965}, typically aimed at PLL operation.  More details of the calculus can be found in the appendix \ref{sec_appendix:phase-slip-dynamics} and for a more detailed study of the diffusion in a tilted periodic potential it is referred to Refs. \cite{Lindner2001,Reimann2001}.  
As mentioned, both drift and diffusion parameters classify the phase slips. Hence, they are further discussed in  section \ref{sec:discussion_PLL}.

\section{ Phase noise PSD of a PLL corrected STNO }
\label{sec:discussion_PLL}

The given results in section \ref{sec_PLL:theory} are quite generally applicable to describe the phase dynamics under noise for any external periodic forcing. 
For PLL operation, 
 the external signal corresponds to a modulation feedback with $n=1/N$. 
The feedback signal in the form of a spin polarized current acts  on the STNO amplitude\cite{Slavin2009} and hence the coupling parameters can be set to $\epsilon_p \neq 0$ and ${\epsilon}_{\phi} = 0$. Note that this situation simplifies the discussion but however, the results remain qualitatively valid even when ${\epsilon}_{\phi} \neq 0$ (see appendix \ref{sec_appendix:renormalization}). 
From eqs. \eqref{eq_theo:mutual_1}, the important PLL bandwidth can be calculated. 
It describes the locking capability defined by the difference between the external frequency $\omega_e$ and the multiple of the natural frequency $n\omega$: 
\begin{align}
\Delta \omega_{BW} = n \nu \epsilon_p a_{\text{ext}} = n\nu \Omega_p \frac{a_{\text{ext}} }{\sqrt{p_0}}  ~~~.  \label{eq_PLL:PLL-bandwidth}
\end{align} 
Interestingly, the PLL bandwidth (eq. (\ref{eq_PLL:PLL-bandwidth})) scales with the nonlinearity factor $\nu$. The latter consequently favors an efficient phase locking through the PLL. 
The coupling strength $\Omega_p$ depends on the interaction mechanism and the STNO's intrinsic parameters. 
For STNOs, the PLL feedback mechanism is usually a direct correction current, for which $\Omega_p a_{\text{ext}}\sim \left. d\Gamma(I)/dI\right|_{I_{dc}} I_{\text{PLL}}$, or a magnetic correction field generated by a field line.  

\subsection{Phase noise PSD in the phase locked state without phase slips}
\label{sec:discussion_without-phase-slips}

 \begin{figure}[hbt!]
\centering
\vspace{-0.0cm}
\includegraphics[width=0.72\textwidth]{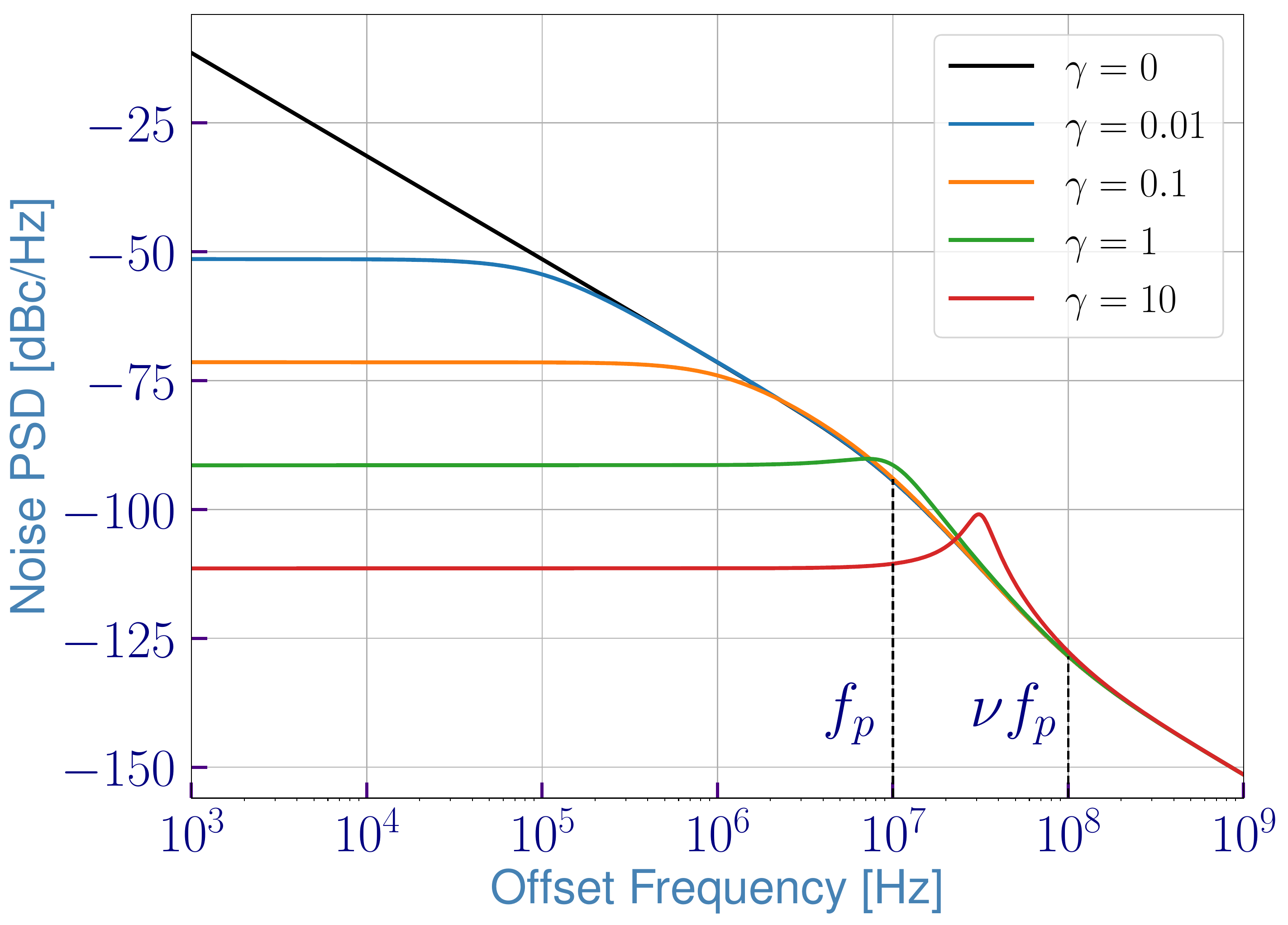}
\caption[Theoretical phase noise PSD for different PLL coupling strengths]{Phase noise PSD for different  parameters of $\gamma$. The black curve represents the free-running STNO. }
\label{fig_PLL:noise-plot-theo_PLL}
\end{figure}

In fig. \ref{fig_PLL:noise-plot-theo_PLL}, we display the evaluated phase noise PSD (eq. (\ref{eq_PLL:PN})) for different values of $\gamma$ corresponding to the ratios of  amplitude coupling $\epsilon_p$, PLL divider rate $n$ and nonlinearity $\nu$ (eq. \eqref{eq:gamma_u}). 
Evaluating eq. \eqref{eq:gamma_u} with the bandwidth, eq. \eqref{eq_PLL:PLL-bandwidth}, the parameter $\gamma$ describes the PLL bandwidth relative to the nonlinear damping rate: $\gamma \sim \Delta \omega_{BW}/(2\Gamma_p)$. 
To compare with the experimental results from section \ref{sec_PLL:STVO}, we set the parameters $f_p=\Gamma_p/\pi = 10\,$MHz, $\nu = 10$ and $\text{FWHM}=2\Delta \omega_0 \left(1 + \nu^2 \right)   =450\,$kHz fixed. %
The results reproduce well the different features observed in the experiments (sec. \ref{sec_PLL:STVO}), such as the flattening of phase noise at low offset frequencies and a resonance peak around the cutoff frequency. 
From eq. (\ref{eq_PLL:PN}), the  offset frequency limit values in the graph \ref{fig_PLL:noise-plot-theo_PLL} can be determined:
\begin{subequations}
\begin{gather}
\lim_{\omega \rightarrow \infty} S_{\delta \psi^2} (\omega) = \frac{\sigma_0^2}{\omega^2} \label{eq_PLL:high-frequ-limit}  \\[1ex]
\lim_{\omega \rightarrow 0} S_{\delta \psi^2} (\omega)  \nonumber \\[1ex] 
{=} \frac{\sigma_0^2 (1+\nu^2) }{   {a_{\text{ext}}^2 n^2 \epsilon_p^2 \nu^2}   -  {\Delta\omega^2}     }  =  \frac{\sigma_0^2 (1+\nu^2) }{   \Delta\omega_{BW}^2   -  {\Delta\omega^2}     } ~~~.  \label{eq_PLL:low-frequ-limit}
\end{gather}  
\end{subequations}
In the high offset frequency limit, eq. \eqref{eq_PLL:high-frequ-limit},  the phase noise PSD  of the PLL-corrected signal joins the one of a free running oscillator (black curve in fig. \ref{fig_PLL:noise-plot-theo_PLL}). \\
In the low offset frequency limit, eq. \eqref{eq_PLL:low-frequ-limit}, $S_{\delta \psi^2}$ is independent of $\omega$ and thus adopts a constant value. 
Its level is inversely proportional to the PLL bandwidth $\Delta \omega_{BW}$. Through $\Delta \omega_{BW}$ (see eq. \eqref{eq_PLL:PLL-bandwidth}), the parameters $n$ and $\epsilon_p$ can lower the low offset frequency noise plateau for increasing values; the dependence on $\nu$ is more complicated and further discussed in sec. \ref{sec:PLL_parameter-space}. 
Eq. \ref{eq_PLL:low-frequ-limit} also shows that the phase noise level depends on the frequency detuning $\Delta \omega$. 
Large detuning increases the phase noise level until at $\Delta \omega = \Delta \omega_{BW}$, the detuning reaches its maximum in the synchronization bandwidth and the phase noise diverges.

\subsubsection*{Frequency cutoff and resonance in the phase noise PSD}

The noise PSD representation, eq. (\ref{eq_PLL:PN}), can  be interpreted as a filter with transfer function $H(f)$, correlated with the PSD through: $S_Y(f) = \left| H(f) \right|^2 S_X (f)$, with $X$ the input and $Y$ the output signal (here, this is the thermal white noise process and the phase noise, resp.).  
The numerator in (\ref{eq_PLL:PN}) yields a cutting frequency of 
\begin{align*}
\omega_{cut} = 2\Gamma_p \sqrt{1+\nu^2} ~~~,
\end{align*}
\hspace{0.25cm} which arises from the inverse first order lowpass filter of the numerator in eq. (\ref{eq_PLL:PN}). For large nonlinear parameters $\nu>1$, it can be written in terms of the real frequency $f_{cut} \approx \nu f_p$,  also plotted in fig. \ref{fig_PLL:noise-plot-theo_PLL}. 

The denominator in  eq. (\ref{eq_PLL:PN}) represents a low pass filter of 2nd order\footnote{System known as $PT_2$ system, whose parameters can be identified through the amplitude representation $\left| H(i\omega)  \right| \sim 1/\sqrt{\left( 1- \left( \nicefrac{\omega}{\omega_0} \right)^2    \right)^2 + \left( 2\mathcal{D} \cdot \nicefrac{\omega}{\omega_0}  \right) ^2 }$ with filter parameters damping $\mathcal{D}$, quality factor $Q=1/(2\mathcal{D})$ and natural, undamped frequency $\omega_0$ \cite{Schulz-Regelungstechnik}} with a natural frequency $\omega_n = 2\Gamma_p (1+ u) \sqrt{ \gamma}$ and quality factor $Q=\sqrt{ \gamma}$. For frequencies higher than $\omega_n$, the system decreases by $1/f^4$. 
Subsequently determining the filter frequency taking the damping $\mathcal{D}=1/(2Q)$ into account gives:
\begin{align}
\omega_d = \omega_n \sqrt{1-2\mathcal{D}^2} = 2\Gamma_p(1+ u) \sqrt{ \gamma - \frac{1}{2}}  ~~~.  \label{eq_PLL:2nd-order-filter_damped-frequency}
\end{align}
It allows evaluating the filter stability, similar to a standard harmonic oscillator as can be also determined by the eigenvalues in eq. \eqref{eq_PLL:PN}. 
For values $\gamma < 1/2 $, the filter is overdamped and strictly decreases with the frequency. For  $\gamma>1/2$, the filter is underdamped and peaks at $\omega_d$, also represented in fig. \ref{fig_PLL:noise-plot-theo_PLL} for higher values of $\gamma$. 
Comparing the maximum filter value at $\omega_d$ with the low frequency limit (\ref{eq_PLL:low-frequ-limit}), the amplitude of the overshoot peak is expressed as: 
\begin{align*}
S(\omega_d) = \frac{4\gamma^2}{4\gamma -1}  ~ {S(\omega \rightarrow 0)} ~~~.
\end{align*}
 Thus, the overshoot becomes more pronounced at higher values of $\gamma$, which is  related to the coupling $\epsilon_p$, divider ratio $n$ and nonlinearity parameter $\nu$ (eq. \eqref{eq:gamma_u}), or expressed differently as $\gamma \sim \Delta \omega_{BW}/(2\Gamma_p)$.  
 This is discussed experimentally (sec. \ref{sec_PLL:STVO}) and theoretically shown in fig. \ref{fig_PLL:noise-plot-theo_PLL}.

\subsection{Parameters of the phase slip dynamics}
\label{sec_PLL:discussion}

In general, the occurrence of phase slips should be reduced for efficient PLL operation and hence, the effective drift $v_e$ and diffusion $D_e$ in the synchronization potential $V(\psi)$ minimized.

\begin{figure}[bth!]
  \centering  %
 
 \subfloat[ \label{fig_PLL:drift_diffusion_vs_F-2}] 
  { %
  \includegraphics[width=0.685\textwidth]{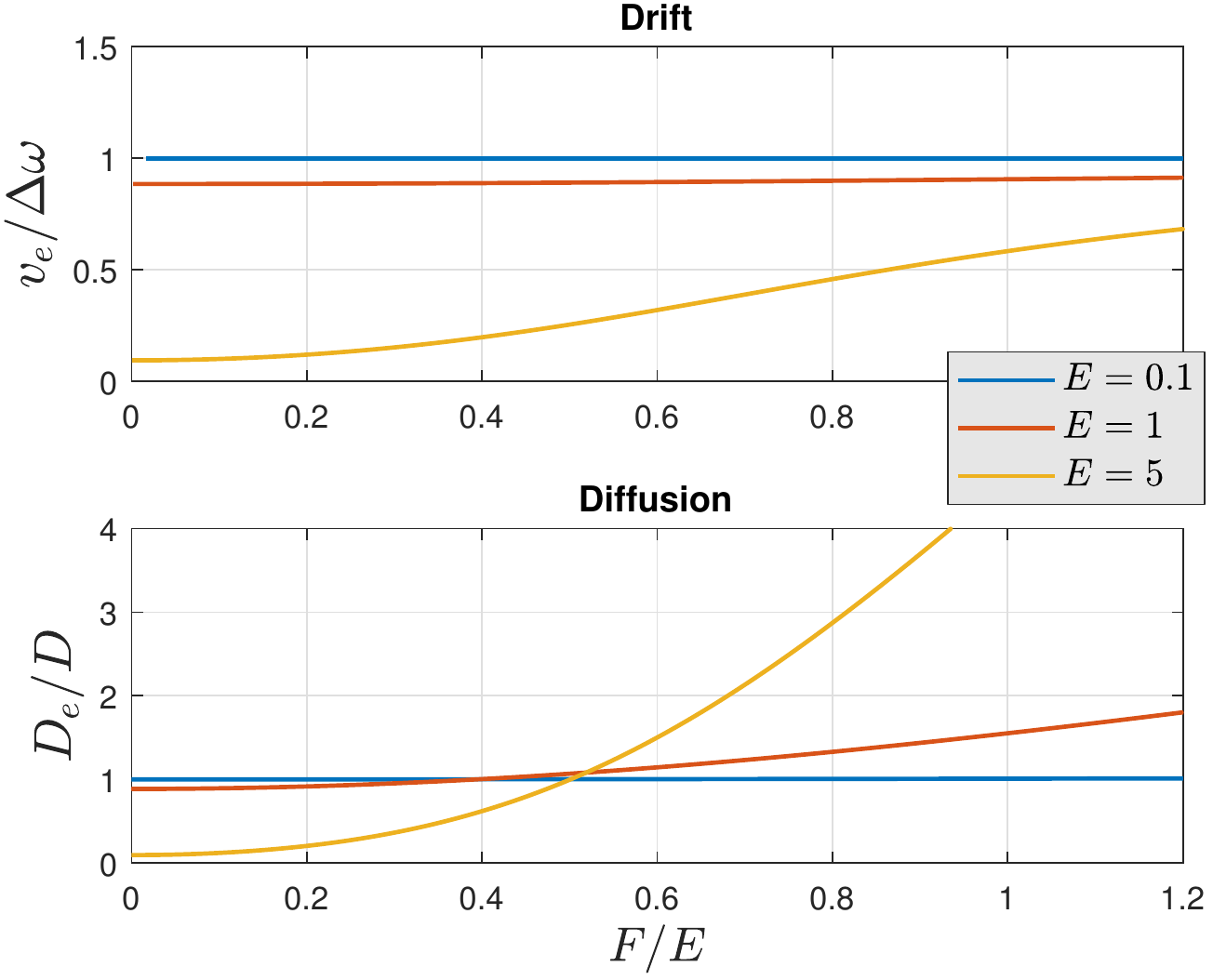}  } 
  
   \subfloat[ \label{fig_PLL:diffusion_vs_E-1}]
  {  %
  \includegraphics[width=0.691\textwidth]{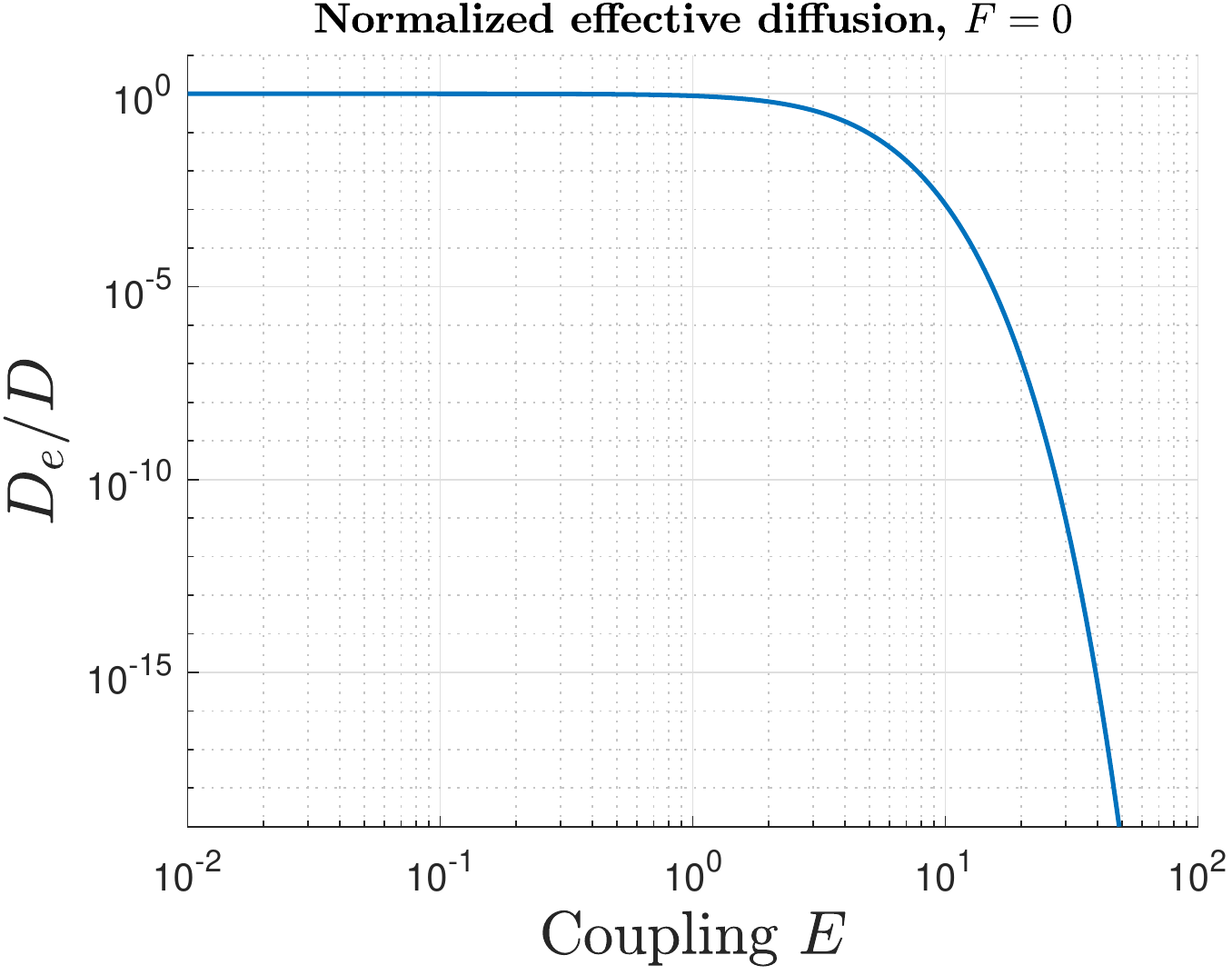}  }

  \caption[Drift and diffusion vs. coupling strength and frequency mismatch]{Coefficients determining the phase slip dynamics in the periodic potential $V$: (a) Drift and diffusion coefficients $v_e$ and $D_e$  as a function of the normalized frequency mismatch $F$ at different coupling strength values $E$. (b) Normalized effective diffusion for $F=0$ as a function of the coupling $E$. 
}
  \label{fig_PLL:phase-slips_drift_and_diffusion}
\end{figure}

Thus, we characterize in fig. \ref{fig_PLL:phase-slips_drift_and_diffusion}  their dependence on the PLL and STNO parameters. 
Therefore,  the normalized coupling $E =  2a_{\text{ext}} \nu \epsilon_p /(nD)$, defining the relative potential barrier height between phase slips,  and frequency mismatch $F=2 \Delta\omega/(nD)$ are introduced. 
Fig. \ref{fig_PLL:drift_diffusion_vs_F-2} reveals a strong influence on the frequency mismatch: For small normalized detuning $F$, we observe that for higher coupling $E$ the effective diffusion $D_e$ and also the effective drift coefficient $v_e$ can be minimized, hence the phase slips suppressed resulting in a locking of the STNO. 
However, the situation changes for higher detuning values $F$: The drift coefficient converges to the detuning $\Delta \omega$ and the effective diffusion $D_e$ significantly increases. 
Even for increasing coupling $E$, the effective diffusion concurrently increases further if the target frequency is too far from the free-running frequency. 
This behaviour of $v_e$ and $D_e$ with  detuning $F$ and coupling $E$ enables us to understand the difference in the two measured blue curves of coupling value $\epsilon = 0.152$ and $\epsilon = 0.153$ in fig. \ref{fig_PLL:PLL_20_PN_conditions}: We see that despite a larger coupling, the noise level and as well the phase slip dynamics can be less favorable. This is because of the larger frequency mismatch of the specific curve, which, in the picture of fig. \ref{fig_PLL:drift_diffusion_vs_F-2} increases the effective diffusion constant $D_e$. 
It is to be noted that $D_e$ can even exceed the intrinsic phase diffusion $D$ at  large detuning (fig. \ref{fig_PLL:drift_diffusion_vs_F-2}). 
In case of a sufficiently low detuning, a higher coupling results in a  decrease  of the effective diffusion as it is depicted in graph \ref{fig_PLL:diffusion_vs_E-1}, in excellent agreement with the presented measurement data in section \ref{sec_PLL:STVO}. 

\subsubsection*{Phase slip limit considerations}

On long time scales (see eq. (\ref{eq_PLL:drift-diffusion_time-limes})) it is: 
 $\braket{\dot{\psi}} = v_e$ and the diffusion describes a Gaussian noise process with variance $\Delta \psi^2 = \braket{ \psi^2 } - \braket{\psi}^2 = 2 D_e t$, i.e. a $1/f^2$ phase noise PSD characteristics\cite{Wittrock2020-SciRep}.  
If phase slips occur, the STNO is never perfectly locked in terms of a noise plateau at low offset frequencies. 
Thus,  the phase noise PSD always readopts a $1/f^2$ shape with $S_{\delta\psi^2}(\omega)= D_e/\omega^2$ at low offset frequencies below a characteristic frequency at which phase slips occur. 

At high diffusion $D\rightarrow \infty$ (high temperature),  the normalized magnitudes $F$ \& $E$ tend to zero implying the modified Bessel function to equal one: $D\rightarrow \infty$ $\Rightarrow$ $F$, $E\rightarrow 0$ $\Rightarrow$ $I_{iF/2} \left( E/2 \right) \rightarrow 1$. 
Consequently, drift and diffusion (eqs. \eqref{eq_PLL:phase-drift_equation} \& \eqref{eq:diffusion_coeff}) correspond to the free-running oscillation:  $v_e \rightarrow \Delta \omega$ \& $D_e \rightarrow D$ and the periodic potential is negligible.  

For large coupling $E\gg 1$, the situation depends on the detuning $F$, as discussed above. 
At simultaneously small detuning $F\ll 1$, it follows $ \left| I_{iF/2} \left( E/2 \right) \right|^{-2} \rightarrow \pi E e^{-E}$. 
As a consequence, it means that at small detuning, phase drift $v_e$ and as well diffusion $D_e$ exponentially decrease with the coupling strength $E$, likewise decreasing the phase slip probability.  
This again agrees well with the performed measurements presented in section \ref{sec_PLL:STVO}.

 \subsection{Discussion: Phase noise PSD including phase slips: PLL performance and parameter space}
\label{sec:PLL_parameter-space}

\begin{figure*}[bth!]
  \centering  %
  \subfloat[ \label{fig_PLL:noise_different_coupling_with_phase-slips-1}]
  {  %
  \includegraphics[width=0.315\textwidth]{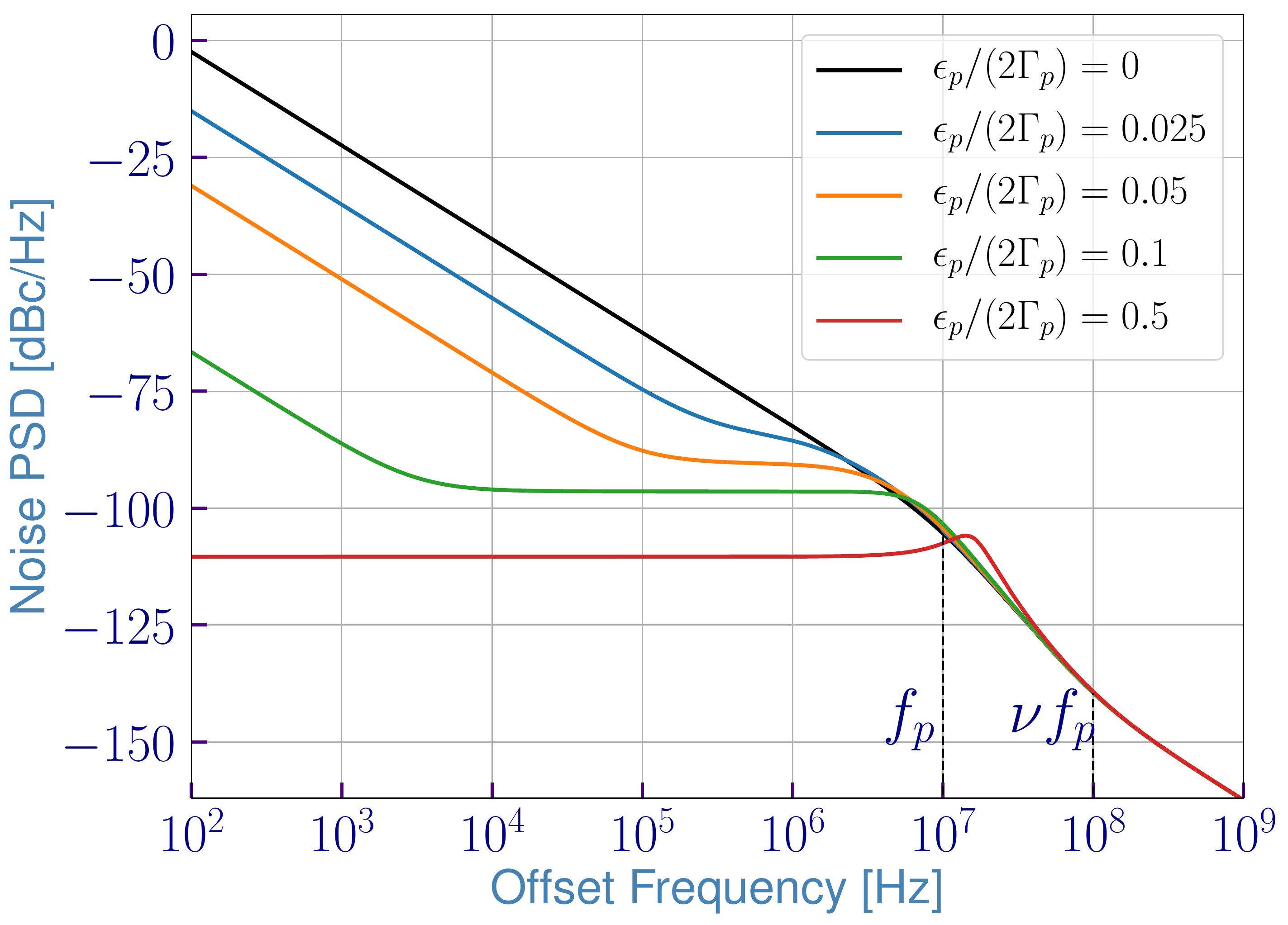}  }   
 \subfloat[ \label{fig_PLL:noise_different_params_n_nu-2}] 
  { %
  \includegraphics[width=0.31\textwidth]{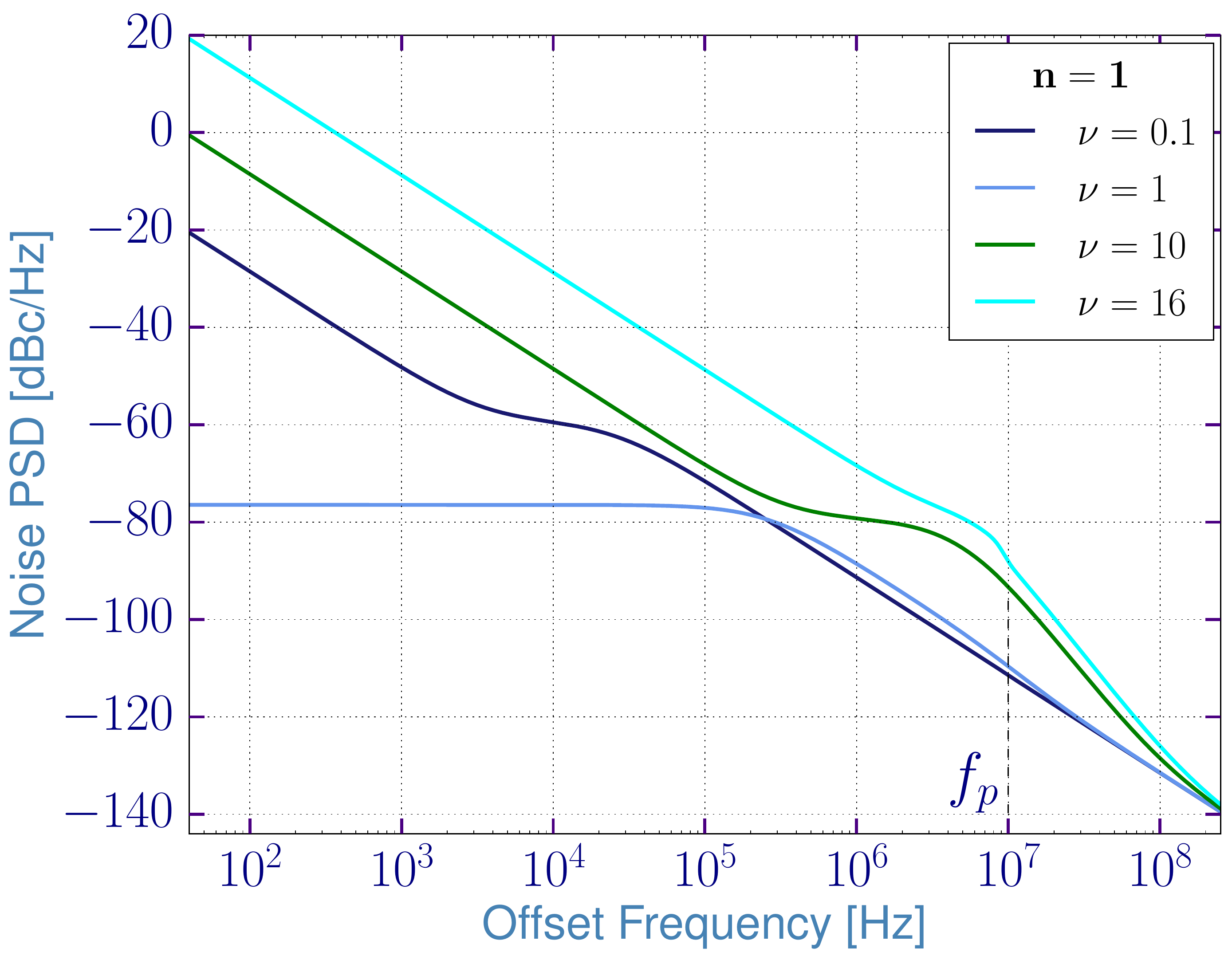}  } 
  \subfloat[ \label{fig_PLL:noise_different_params_n-2}] 
  { %
  \includegraphics[width=0.31\textwidth]{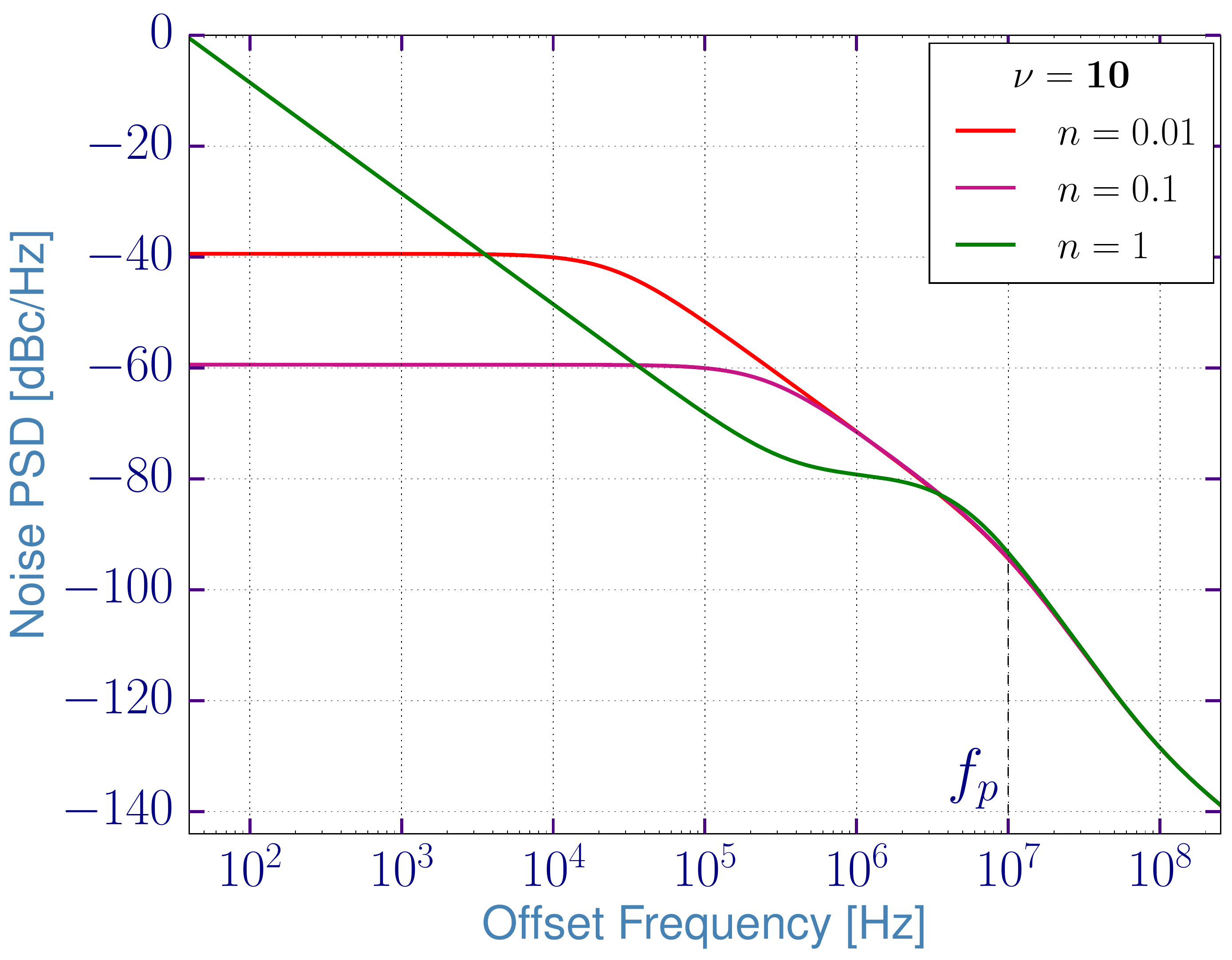}  } 
  \caption[Drift and diffusion vs. coupling strength and frequency mismatch]{Phase noise (PN) PSDs covering the parameter space of $\nu$, $n$ and $\epsilon_p$. (a) PN for different coupling coefficients $\epsilon_p$. (b) PN for different parameter values $\nu$ at $n=\text{const}=1$. (c) PN for different parameter values $n$ at $\nu=\text{const}=10$.  The green curve in (b) and (c) is the same one. The following constant values are chosen: $F=0$, $f_p=1\cdot 10^7\,$Hz, $\text{FWHM}=450\,\text{kHz}=4\pi (1+\nu^2)\Delta f_0 $, and, if not explicitely indicated differently, $\nu=10$, $\epsilon_p=0.025$, $n=1$.
}
  \label{fig_PLL:noise_different_parameters_with_phase-slips}
\end{figure*}

In fig. \ref{fig_PLL:noise_different_parameters_with_phase-slips}, we illustrate more in detail  the dependence of the phase noise PSD on the system parameters $\epsilon_p$, $\nu$ and $n$. 
The contribution due to phase slips $S_{\delta\psi^2}(\omega)= D_e/\omega^2$ can be summed to the phase noise PSD at efficient locking (eq. \eqref{eq_PLL:PN}). 
In fig. \ref{fig_PLL:noise_different_coupling_with_phase-slips-1}, the lowered noise plateau at low offsets with increasing coupling is observed for constant values of $\nu$ and $n$. For the highest shown coupling value, a resonance bump is seen slightly above $f_p$, as discussed in sec. \ref{sec:discussion_without-phase-slips}. 
 Furthermore, the discussed exponential decrease $\sim \exp(-E)$ of the phase slip effective diffusion $D_e$ with the coupling strength can be observed (here, $F=0$), decreasing the phase slip phase noise contribution for increasing coupling. 
Fig. \ref{fig_PLL:noise_different_params_n_nu-2} summarizes the complex dependence of the noise PSD on the  STNO  nonlinearity parameter $\nu$ for constant coupling $\epsilon_p/(2\Gamma_p)=0.025$ and $n=1$.  
As it is derived in sec. \ref{sec_PLL:theory} (see eq. (\ref{eq_PLL:PLL-bandwidth})), we observe that the PLL bandwidth is enhanced by the oscillator's nonlinearity $\nu$.  
However,  $\nu$   also strongly broadens the STNO linewidth and enhances the diffusion by $(1+\nu^2)$ \cite{Tiberkevich2008,Slavin2009}, which can in fig. \ref{fig_PLL:noise_different_params_n_nu-2} mainly be recognized at higher offsets. 
Moreover, the noise overshot at the resonance frequency of the 2nd order filter transfer function  (see eq. (\ref{eq_PLL:2nd-order-filter_damped-frequency})) is likewise proportional to $\nu$ and  narrows the PLL performance. 
Thus, the filter overshoot at $\omega_d$ close to the PLL bandwidth becomes more pronounced with the nonlinearity, also recognized in fig.  \ref{fig_PLL:noise_different_params_n_nu-2}. 
Furthermore taking the phase jump dynamics into account, the normalized coupling strength $E$ is proportional to $\nu$ and antiproportional to $nD$. On the contrary, it is $D\sim (1+\nu^2)$ and thus the nonlinearity lowers the PLL performance  in terms of phase slip dynamics, what is perfectly retrieved in the curves of fig. \ref{fig_PLL:noise_different_params_n_nu-2} in which the best noise characteristics are found for $\nu=1$.  

As a last parameter, the divider ratio $n$ is to be considered. Its influence on the phase noise is shown in fig. \ref{fig_PLL:noise_different_params_n-2} for $\nu = \text{const}=10$:  A high value leads to a lower noise plateau at low frequencies but likewise also increases the   resonant 2nd order filter response (see sec. \ref{sec:discussion_PLL}). 
Again taking the phase slip dynamics into account, $n$ decreases the coupling strength $E\sim (nD)^{-1}$ and phase slips are more likely to occur at high $n$, especially recognized for $n=1$ in fig.  \ref{fig_PLL:noise_different_params_n-2}.

The discussed dependences are in complete agreement with the presented measurements in section \ref{sec_PLL:STVO}. 
The discussion highlights that an adequate trade-off between the described mechanisms and parameter dependences must always be found for an efficient noise suppression by the PLL.

\section{Conclusion}
\label{sec_PLL:conclusion}

We present the experimental implementation of two different STNOs -- a STVO  and a uniform STNO (section \ref{sec_PLL:STVO}) -- into an on-chip integrated PLL developed for this purpose. 
We find an efficient phase noise reduction of $50\,$dB @ $10\,$kHz for the vortex and of $20\,$dB @ $100\,$kHz offset for the uniform STNO with $f_c \approx 340\,$MHz and $f_c \approx 4.8\,$GHz, respectively.  
Furthermore, we thoroughly analyze the PLL performance for different control parameters and find rather complex dependences due to the intrinsic large nonlinearity of our oscillators. 
The occurrence of phase slips, mainly caused by the large intrinsic noise in STNOs, is identified as the main drawback for the exploitation of the PLL system. 

In complement to the experimental results, we provide a complete theoretical framework to analyze the performance of the PLL system upon operation with a nonlinear spin torque nano oscillator. 
This model excellently describes the system's parameter characteristics and reveals the physics of the low offset frequency noise due to phase slip dynamics. 
Note that in general, this theoretical development is not restricted to only the case of a PLL but the approach can easily be employed for any external signal applied to the STNO. 
It is found that the dependence of the noise PSD on the PLL and STNO parameters is complex and its optimization requires a trade-off between all of those, implying a prior analysis of the parameter space. 
Our approach of a highly configurable PLL chip is a perfect basis to handle the system complexity since the important parameters can be easily adjusted\cite{Kreissig2017-AIP,Kreissig2017-IEEE}. 
Thus, it illustrates a first step towards flexible, integrated and hybrid systems for accurate frequency generation utilizing STNOs.

\section*{Acknowledgment}

S.W. acknowledges financial support from Labex FIRST-TF under contract number ANR-10-LABX-48-01. 
The work is supported by the French ANR project ''SPINNET'' ANR-18-CE24-0012. We acknowledge the Plateforme Technologique Amont (PTA), Grenoble, France for access and support for the nanofabrication

\bibliography{literatur_promo}

\clearpage 
\appendix

\counterwithout{equation}{section}
\addtocounter{equation}{-1}
\counterwithout{figure}{section}
\addtocounter{figure}{-1}

\renewcommand\thefigure{A.\arabic{figure}} 
\setcounter{figure}{0}  
\renewcommand\theequation{A.\arabic{equation}} 
\setcounter{equation}{0}

\section{Experimental details}
\label{sec_appendix:Exp}

\subsection{Samples}

\selectlanguage{english}

The measured vortex based and uniform STNOs have the following structure: BE/AF/SAF/MgO/FL/cap. 
SAF denotes the polarizing layer and consists of a synthetic antiferromagnet (SAF) pinned by an antiferromagnet (AF). 
It has the same structure for both types of devices with 
 \selectlanguage{ngerman} PtMn($20$)/""Co$_{70}$Fe$_{30}$($2$)/""Ru($0.85$)/""Co$_{40}$Fe$_{40}$B$_{20}$($2.2$)/""Co$_{70}$Fe$_{30}$($0.5$) \selectlanguage{english} and the nm layer thickness in brackets. 
 MgO is the tunnel barrier whose thickness and oxidation time is adjusted to yield a nominal resistance area product of $1.45\,\Omega\upmu$m$^2$ ($1.5\,\Omega\upmu$m$^2$) for the vortex (uniform) MTJs. 
 The free layer (FL) is a bilayer of \selectlanguage{ngerman} Co$_{40}$Fe$_{40}$B$_{20}$($t_1$)/""Ta($0.2$)/""Ni$_{80}$Fe$_{20}$($t_2$)
  \selectlanguage{english} with thicknesses $t_1 = 1.5\,$nm ($2\,$nm) and $t_2 = 7\,$nm ($2\,$nm) for the vortex (uniform) MTJs, respectively. 
 The Ta serves as a Boron pump and decouples the CoFeB and NiFe layers to assure a high  tunneling magnetoresistance (TMR). BE denotes the bottom electrode material with \selectlanguage{ngerman} Ta(3)/""CuN(30)/""Ta(5)    and cap is the capping material with  Ta(2(3))/Ru(7)   \selectlanguage{english}  for the vortex (uniform) devices. 
 All MTJ stacks were sputter deposited by INL using a Singulus-Timaris deposition tool onto high resistivity Si substrates with an additional $500\,$nm SiO$_2$ layer. The MTJ stacks are subsequently annealed for $2\,$h at $T = 330\,^{\circ}$C at an applied magnetic field of $1\,$T. They were then patterned at the PTA facilities (Grenoble) by SPINTEC using standard optical and e-beam lithography processes as well as Ar ion etching. The magnetoresistance values (at room temperature) of the STNO nanopillars were around 50\%. Data are presented for vortex (uniform) STNOs with diameter of $D = 300 (80)\,$nm, leading to an emission frequency range of $200$-$400\,$MHz (under an out-of-plane field) for the vortex STNOs and $3$-$6\,$GHz (under in-plane field) for the uniform STNOs.

\subsection{Measurements}
\label{sec_appendix:measurement-setup}

The different STNO devices are directly employed on the presented PLL chip. 
It delivers the dc current in order to operate the STNO at self-oscillations, which are sustained by the resulting spin transfer torque. 
The PLL chip with STNO is subjected to an applied magnetic field that is specified in the related sections. 
The emitted rf  time signals are recorded by a single-shot oscilloscope measurement and the emission spectra are gathered by simultaneous employment of a spectrum analyzer. 
In order to obtain noise data, the measured time signal is processed via the Hilbert transform method \cite{Wittrock2019_PRB,Bianchini2010,Quinsat2010}.

\section{Theory -- efficient PLL phase locking }

\subsection{Power spectral density}
\label{sec_appendix:PSD-calculus}

The formal solution of eq. (\ref{eq_PLL:matrix-SDE}) is: 
 \begin{align*}
 X_t = e^{-At}X_0 + \int_0^t e^{-A (t-s)} \Sigma H_s ds  ~~~,
 \end{align*}
\hspace{0.25cm} and by evaluating  the eigenvalues of $A$, a classical stability analysis can be performed. 
However, to determine the noise PSD of the system, a Fourier transform of system (\ref{eq_PLL:matrix-SDE}) is conducted. Assuming $H_t$ deterministic and continuous for the Fourier transform to be defined, one obtains:
 \begin{align*}
 \hat{X} (\omega) = \left( A+ \mathbbm{1} i\omega \right)^{-1} \Sigma \hat{H}_t (\omega)  ~~~.  
 \end{align*}  %
 With $S_H (\omega)$ the PSD of $H_t$, the noise PSD of $X_t$ is in consequence described by:
  \begin{align*}
  S_X(\omega) = \left( A+ \mathbbm{1} i\omega \right)^{-1} \Sigma S_H (\omega) \Sigma^{\intercal} \left( A^{\intercal} - \mathbbm{1} i\omega \right)^{-1}   ~~~.
  \end{align*}
Here, the white noise processes in $H_t$ are independent and thus its PSD is diagonal with each process' covariance as the entries of $S_H (\omega) = \mathbbm{1}$. 
Thus, one obtains:    
\begin{align*}
S_X(\omega) = \sigma_0^2 (A + \mathbbm{1} i\omega)^{-1} (A^{\intercal} - \mathbbm{1} i\omega)^{-1}  ~~~.
\end{align*}

Taking the matrix $A$ in its general form:
\begin{align*}
A = \begin{pmatrix}
a_{11} & a_{12} \\
a_{21} & a_{22}
\end{pmatrix}  
\end{align*}
\hspace{0.25cm} the inverse matrix can be computed:
\begin{align*}
S_X^{-1} \sigma_0^2 = \left( A^{\intercal} - \mathbbm{1} i\omega \right) \left( A + \mathbbm{1} i\omega \right) \\
= \begin{pmatrix} a_{11}^2 +a_{21}^2   + \omega^2      &    \Lambda^*    \\ 
\Lambda    &   a_{12}^2 + a_{22}^2 + \omega^2       \end{pmatrix}  ~~~,
\end{align*}
\hspace{0.25cm} with $\Lambda = a_{11}a_{12} + a_{21}a_{22} + i \omega (a_{12} - a_{21})$. 
Its determinant is given through the systems' eigenvalues $\lambda$ by:
\begin{align*}
\mathcal{D}= \det\left( S_X^{-1} \sigma_0^2 \right) &= \det\left( A^{\intercal} - \mathbbm{1} i\omega \right) \det\left( A + \mathbbm{1} i\omega \right)   \\
& = (\lambda_1 - i\omega) (\lambda_2 - i\omega)  (\lambda_1 + i\omega) (\lambda_2 + i\omega)\\ 
& = (\lambda_1^2+\omega^2)(\lambda_2^2+\omega^2) ~~~. 
\end{align*}

Now inverting $S_X^{-1} \sigma_0^2$, one obtains for the general form of the noise power spectral density matrix:
  \begin{align}
  S_X(\omega) &= \frac{\sigma_0^2}{\mathcal{D}} \operatorname{adj}\left( S_X^{-1} \sigma_0^2 \right) \nonumber \\
  & = \frac{\sigma_0^2}{\mathcal{D}} \begin{pmatrix}
  a_{12}^2 + a_{22}^2 + \omega^2  &   - \Lambda^* \\
  -\Lambda  &  a_{11}^2 + a_{21}^2 + \omega^2
  \end{pmatrix}   ~~~.  \label{eq_PLL:appendix_complete_PSD_matrix}
  \end{align}

\subsection{Generalization through renormalization of the coupling terms}
\label{sec_appendix:renormalization}

In a general case, the theoretical calculations starting from eqs. (\ref{eq_theo:mutual_1}) might be simplified by the introduction of a complex notation, allowing for the generalization of some of the discussed results when $\epsilon_p , \epsilon_{\phi} \neq 0$:
\begin{align*}
\bar{\epsilon}_p = \epsilon_p e^{i\beta_p} ~, \quad  \bar{\epsilon}_{\phi} = \epsilon_{\phi} e^{i\beta_{\phi}} ~,  \quad    \bar{\epsilon} = \epsilon_p e^{i\beta}  ~~~.   
\end{align*}
Here, the last variable is the new general coupling and defined by:
\begin{align*}
\bar{\epsilon} =  i \bar{\epsilon}_p +  \bar{\epsilon}_{\phi}/\nu  ~~~,
\end{align*}
\hspace{0.25cm} with the norm $\epsilon$ and argument $\beta$ given by:
\begin{align*}
\epsilon &= \sqrt{ \epsilon_p^2 + \epsilon_{\phi}^2/\nu^2 + 2\epsilon_p\epsilon_{\phi} \sin(\beta_{\phi} -\beta_p)/\nu  } \\
\cos(\beta) &= -\epsilon_p/\epsilon \cdot \sin(\beta_p) + \epsilon_{\phi}/(\nu \epsilon) \cos(\beta_{\phi})  \\
\sin(\beta) &= \epsilon_p/\epsilon \cdot \cos(\beta_p) + \epsilon_{\phi}/(\nu\epsilon) \sin(\beta_{\phi}) ~~.
\end{align*}
This gives the following substitutions for a real variable $x$:
\begin{align*}
\epsilon\cos(x-\beta) &= \epsilon_p \sin(x-\beta_p) + \epsilon_{\phi}/\nu \cos(x-\beta_{\phi}) \\
\epsilon\sin(x-\beta) &= -\epsilon_p \cos(x-\beta_p) + \epsilon_{\phi}/\nu \sin(x-\beta_{\phi}) ~~.
\end{align*}
This approach can be used in order to generalize the results that are specifically discussed for the situation of a PLL, such as eqs. \eqref{eq_PLL:PLL-bandwidth} or \eqref{eq_PLL:low-frequ-limit}. 
It allows for a more qualitative discussion, not only limited to the case of a PLL, but indeed, for any applied external synchronization signal. 
For instance, the stochastic Adler equation (eq. \eqref{eq_PLL:Adler_for_PLL}) would become:
\begin{align*}
    \dot{\psi} = \Delta \omega - \nu \epsilon \sin(n\psi -\beta) + \sigma_0 \sqrt{1+\nu^2}\eta ~~~,
\end{align*}
\hspace{0.25cm} which is more easily recognized as the common stochastic Adler equation due to the renormalized coupling term.

\section{Theory -- Phase slip dynamics}
\label{sec_appendix:phase-slip-dynamics}

The dephasing dynamics is represented in fig. \ref{fig_PLL:potential-field-1}. 
It is described as a Brownian motion in a periodic potential field, given by:
\begin{align}
V(\psi)  = -\left[ \vphantom{\frac{\nu \epsilon_p a_{\text{ext}}}{n}} \psi \Delta \omega  \right. &   \left. + \frac{\nu \epsilon_p a_{\text{ext}}}{n} \sin(n \psi - \beta_p ) \right. \nonumber  \\[1ex] 
&  \left.  + \frac{\epsilon_{\phi} a_{\text{ext}}}{n} \cos(n \psi - \beta_{\phi}  ) \right]   \label{eq_PLL:phase-slips_potential-field} ~~~,
\end{align}
\hspace{0.25cm} with diffusion $D=(1+\nu^2)\sigma_0^2/2$. 
Local minima of $V$ are found at $\psi=\psi_{eq}+2\pi k/n, k\in \mathbb{Z}$. 
Phase slips are induced by thermal fluctuations and described by jumps between the minima in $V$. 
The drift is determined by $\Delta\omega$ with favored drift direction set by the sign of $\Delta \omega$. 

The fluctuation dynamics can now be described through   the probability density $\mathscr{P}(\psi,t)$ of the dephasing value $\psi$ at time $t$ within the Fokker-Planck formalism \cite{Moss1989,Risken1989}: 
\begin{align}
\frac{\partial \mathscr{P}}{\partial t} = \frac{\partial }{\partial \psi} \left[ V' \mathscr{P} \right] + D \frac{\partial^2 \mathscr{P}}{\partial \psi^2} ~~~,  \label{eq_PLL:Fokker-Planck}
\end{align}
\hspace{0.25cm} with $V'=dV/d\psi$. 
Subsequently, the continuity equation is described by:
\begin{align*}
\frac{\partial \mathscr{P}}{\partial t} \left( \psi , t \right) = - \frac{\partial j_{\mathscr{P}} }{\partial \psi} \left( \psi , t\right) ~~~,
\end{align*}
\hspace{0.25cm} with the probability current 
\begin{align}
j_{\mathscr{P}} (\psi,t) = - V' \mathscr{P} - D ~ \partial \mathscr{P}/\partial \psi  ~~~.  \label{eq_PLL:probability-current}
\end{align} 
For $\mathscr{P}_0$ the stationary probability, $j_{\mathscr{P}}$ is constant and equals the drift $v_e$. 
It is in a limit value examination:
\begin{align*}
\lim_{t\rightarrow \infty} \mathscr{P} (\psi,t) = \mathscr{P}_0(\psi) \frac{Z}{\sqrt{4\pi D_e t}} e^{- \frac{\left( \psi - v_e t \right)^2}{4D_e t}}  ~~~,
\end{align*}
\hspace{0.25cm} with $Z$ a normalization constant. 
The solution for the probability density $\mathscr{P}_0$ is $2\pi/n$ periodic and can be evaluated solving eq. (\ref{eq_PLL:probability-current}):
\begin{align}
\mathscr{P}_0 (\psi) &= -\left( \int\limits_{-\infty}^{\psi} j_{\mathscr{P}} ~ e^{V/D} d\psi' \right) e^{-V/D} \nonumber \\ 
&= - \frac{v_e}{D} \frac{e^{\pi F}}{1-e^{\pi F}} \underbrace{  \int\limits_{\psi}^{\psi + \frac{2\pi}{n}} e^{\frac{ V(\psi') -V(\psi) }{D}} d\psi' }_{=: \mathcal{I}_-(\psi)} ~~~. \label{eq_PLL:phase-slips_stationary_probability}
\end{align}

We introduce the normalized frequency mismatch $F=2\Delta\omega/(nD)$ and define the integral $\mathcal{I}_{\mp}(\psi)$ with index $\mp$ denoting the equivalent integration variables $[\psi \rightarrow \psi+2\pi/n]$, corresponding to a \textit{"-"} sign in the argument of the exponential, and $[\psi - 2\pi/n \rightarrow \psi]$, corresponding to a \textit{"+"} sign in the argument of the exponential: $V(\psi') \mp V(\psi)$ . 
We took advantage of the periodicity of $V$, disrespecting the linear drift, i.e. $V(\psi - 2\pi k/n) = V(\psi) + 2\pi k/n \cdot \Delta\omega$. Thus, the integral on the left-hand side of eq. (\ref{eq_PLL:phase-slips_stationary_probability})  can be evaluated in a geometrical series\cite{Stratonovich1965,Reimann2002}: 
\begin{align*}
\int\limits_{-\infty}^{\psi} e^{V(\psi)/D} d\psi = \sum_{l=0}^{\infty} \int\limits_{\psi}^{\psi + \frac{2\pi}{n}}  e^{\frac{V(\psi - \frac{2\pi}{n} )}{D}} d\psi \\
 = \frac{1}{1-e^{\pi F}} \int_{\psi}^{\psi + \frac{2\pi}{n}} e^{V(\psi)/D} d\psi  ~~~.
\end{align*}

From eq. (\ref{eq_PLL:phase-slips_stationary_probability}), drift $v_e$ and diffusion $D_e$ coefficients, which basically determine the phase slip dynamics, can be evaluated.

\subsubsection{Drift coefficient $v_e$}

The drift coefficient $v_e$ is identified through the normalization of  $\mathscr{P}_0$ \cite{Stratonovich1965,Reimann2002} in eq.  (\ref{eq_PLL:phase-slips_stationary_probability}):
\begin{align*}
\frac{n}{2\pi} \int_0^{\frac{2\pi}{n}} \mathscr{P}_0 (\psi) d\psi = 1  \\[1.6ex]
\Rightarrow  v_e = \frac{D\left( 1- e^{-\pi F}   \right)}{ \frac{n}{2\pi} \int\limits_0^{\frac{2\pi}{n}}  \mathcal{I}_-(\psi) d\psi } ~~~.
\end{align*}
The integral in the denominator was evaluated by Stratonovich\cite{Stratonovich1965}, who finds:
\begin{align}
v_e = \frac{nD}{\pi} \sinh\left( \frac{\pi F}{2} \right) \left| I_{iF/2} \left( E/2 \right)   \right|^{-2}   ~~~,  \label{eq_PLL_appendix:phase-drift_equation}
\end{align}
\hspace{0.25cm} with $I_{ia}(x)$ the modified Bessel function \cite{Bronstein}, $(a,x)\in \mathbb{R}$, and $E = \left[ 2a_{\text{ext}} (\nu \epsilon_p + \epsilon_{\phi})\right]/(nD)$ a normalized coupling.

\subsubsection{Diffusion coefficient $D_e$}

We treat the diffusion constant $D_e$ using the approach presented by  Stratonovich\cite{Stratonovich1965} assuming a small drift (frequency detuning $\Delta \omega$ much smaller than the periodic coupling terms in (\ref{eq_PLL:phase-slips_potential-field})). 
Then, the probability current $v_e$ for the stationary solution $\mathscr{P}_0$ can be treated as a positive and a negative current $j_{\mathscr{P},\pm }$: 
\begin{align*}
v_e = j_{\mathscr{P},+} - j_{\mathscr{P},-}  \quad \text{ with } \quad \frac{j_{\mathscr{P},+}}{j_{\mathscr{P},-} } = e^{-\pi F} ~~~.
\end{align*} 
The effective diffusion is then determined by the relative diffusion through the energy barriers of potential difference $\pi F$ on both sides:
\begin{align*}
D_e = \frac{\pi}{n} \left( j_{\mathscr{P},+} + j_{\mathscr{P},-}     \right) = \frac{\pi}{n} ~ \frac{e^{-\pi F} + 1}{e^{-\pi F} - 1} ~ v_e ~~~,
\end{align*}
\hspace{0.25cm} and is therefore calculated to:
\begin{align}
D_e = D \cosh\left(  \pi F/2       \right)  \left| I_{iF/2} \left( E/2 \right)   \right|^{-2}  ~~~. \label{eq_appendix:diffusion_coeff}
\end{align}
Note that in general, the evaluation of $D_e$ is rather complex. 
However, the derived relation is a good approximation for small normalized frequency detuning $F=2 \Delta\omega/(nD) \ll 1$ \cite{Stratonovich1965}, typically aimed at PLL operation.  A more detailed study of the diffusion in a tilted periodic potential can be found in Refs. \cite{Lindner2001,Reimann2001}.

\end{document}